\pdfoutput=1

\documentclass[sigconf,none,authorversion]{acmart}

\AtBeginDocument{%
  \providecommand\BibTeX{{%
    \normalfont B\kern-0.5em{\scshape i\kern-0.25em b}\kern-0.8em\TeX}}}

\newtheorem{dfn}{Definition}

\usepackage{setspace}
\usepackage{multirow}
\usepackage{booktabs}
\usepackage{colortbl}
\usepackage{paralist}
\usepackage{amsthm}
\usepackage{color}
\usepackage{vcell}
\usepackage[flushleft]{threeparttable}
\usepackage{adjustbox}
\usepackage{rotating}
\usepackage{lscape}
\usepackage{multirow}
\usepackage{afterpage}
\usepackage{longtable}

\begin{document}


\title{Literature Review to Collect Conceptual Variables of Scenario Methods for Establishing a Conceptual Scenario Framework}

\author{Young-Min Baek}
\email{ymbaek@se.kaist.ac.kr}
\affiliation{%
  \institution{Korea Advanced Institute of Science and Technology (KAIST)}
  \city{Daejeon}
  \country{Republic of Korea}
}

\author{Esther Cho}
\email{esthercho@se.kaist.ac.kr}
\affiliation{%
  \institution{Korea Advanced Institute of Science and Technology (KAIST)}
  \city{Daejeon}
  \country{Republic of Korea}
}

\author{Donghwan Shin}
\email{donghwan.shin@uni.lu}
\affiliation{%
  \institution{University of Luxembourg}
  \city{Esch-sur-Alzette}
  \country{Luxembourg}
}

\author{Doo-Hwan Bae}
\email{bae@se.kaist.ac.kr}
\affiliation{%
  \institution{Korea Advanced Institute of Science and Technology (KAIST)}
  \city{Daejeon}
  \country{Republic of Korea}
}

\renewcommand{\shortauthors}{Y. M. Baek, et al.}

\newcommand{\etal}{\textit{et al.}}
\newcommand{\ym}[1]{\color{teal}\texttt{/*\textbf{$@$YM}: {#1}*/}\color{black}}
\newcommand{\DS}[1]{{\color{red} [DS: #1]}}
\newcommand{\RTBR}[1]{{\color{blue} \small\textsf{\textbf{[\textit{READY TO BE REVIEWED}]\\}}}}
\newcommand{\pp}{\color{violet}(\textbf{NEED PARAPHRASING}) \color{black}}

\newenvironment{new_itemize}{
\begin{itemize}
  \setlength{\itemsep}{1pt}
  \setlength{\parskip}{0pt}
  \setlength{\parsep}{0pt}
}{\end{itemize}}

\newenvironment{new_enumerate}{
\begin{enumerate}
  \setlength{\itemsep}{1pt}
  \setlength{\parskip}{0pt}
  \setlength{\parsep}{0pt}
}{\end{enumerate}}

\newenvironment{formal}{%
  \def\FrameCommand{%
    \hspace{1pt}%
    {\color{black}\vrule width 2pt}%
    {\color{formalshade}\vrule width 4pt}%
    \colorbox{formalshade}%
  }%
  \MakeFramed{\advance\hsize-\width\FrameRestore}%
  \noindent\hspace{-4.55pt}%
  \begin{adjustwidth}{}{7pt}%
  \vspace{2pt}\vspace{2pt}%
}
{%
  \vspace{2pt}\end{adjustwidth}\endMakeFramed%
}

\begin{abstract}

Over recent decades, scenarios and scenario-based software/system engineering have been actively employed as essential tools to handle intricate problems, validate requirements, and support stakeholders' communication. However, despite the widespread use of scenarios, there have been several challenges for engineers to more willingly utilize scenario-based engineering approaches (i.e., scenario methods) in their projects. First, the term scenario has numerous published definitions, thus lacking in a well-established shared understanding of scenarios and scenario methods. Second, the conceptual basis for engineers developing or employing scenarios is missing.
To establish shared understanding and to find common denominators of scenario methods, this study leverages well-defined metamodeling and conceptualization that systematically investigate the concepts under analysis and define core entities and their relations. By conducting a semi-systematic literature review, conceptual variables are collected and conceptualized as a conceptual meta-model. As a result, this study introduces scenario variables (SVs) that represent constructs/semantics of scenario descriptions, according to 4 levels of constructs of a scenario method. To evaluate the comprehensibility and applicability of the defined variables, we analyze five existing scenario methods and their instances in automated driving system (ADS) domains. The results showed that our conceptual model and its constituent scenario variables adequately support the understanding of a scenario method and provide a means for comparative analysis between different scenario methods. 

\end{abstract}


\maketitle
\pagestyle{plain}

\section{Introduction}
\label{sec_intro}

Over recent decades, \textit{scenarios} have been considered a familiar technique in software and systems engineering fields, as artifacts or tools/techniques. A scenario, which is originally derived from the Latin word \textit{scena}, meaning ``\textit{scene}'', typically describes something that might happen or is expected to happen and deals with intricate problems and specifications during a development process~\cite{borjeson2006scenario_types, leva2020business_process}. Scenarios also have become frequently-used means to capture and communicate specifications for better understanding among diverse stakeholders from different backgrounds~\cite{greenyer2017scenariotools}. Unlike conventional specifications that strictly describe a system, scenarios are able to provide more readable stories involving imaginable contexts and support generation of executable inputs and exceptions under potential execution/runtime environment. 

Aided by the scenario's versatility, scenario-based engineering approaches, called \textit{scenario methods}, have been actively employed as essential tools for a variety of engineering purposes (e.g., scenario-based analysis \& design~\cite{godet2000art_of_scenario, greenyer2017scenariotools, kurakawa2004scenariodrivendesign}, simulation~\cite{jafer2017tacklingcomplexity, siegfried2012scenarios_in_military, tomizawa2007automated_scenario_generation}, testing~\cite{khastgir2021systems_testscenarios, heinz2017track_scenario_based}). Compared to engineering that does not utilize scenarios, scenario methods facilitate more effective communication and decision making process due to the comprehensible nature of scenarios. Because scenarios have the capability to capture (and lucidly explain) specifications and ease inconsistencies during the development process, they have been used as a medium or a proxy for making the specification more intuitively readable and understandable. By providing a credible and coherent story that involves future possibilities and their contextual information based on hypotheses, scenarios are able to flexibly and agilely provide links between different levels of specifications, such as requirements, design artifacts, and the executables. 

Despite the widespread use of such scenario methods, several challenges exist for scenario engineers to practically apply the scenario method for their development projects. Major reasons disturbing the application of scenario methods can be summarized as (i) the lack of shared understanding of scenarios, and (ii) the lack of a conceptual basis or framework for analyzing and developing a scenario method. As a consequence of these issues, many researchers apply ill-structured scenarios in their own methods or use modeling or specification methods that do not fit the purpose adequately. 

First, due to the lack of a well-established understanding of scenarios, scenario engineers are still making use of scenarios that were developed in an incoherent, ill-structured, casual/informal, or ad-hoc ways. Depending on a target of a scenario method and how engineers see the target problem, scenarios and their semantics can be defined and classified in various ways. In particular, under a growing tendency to employ quantitative scenarios (e.g., scenario-based statistical analysis of risks) rather than qualitative ones (e.g., downscaled scenarios assuming that trends are similar across scales)~\cite{unep2016scenario_development}, engineers need to utilize the scenario methods in the right context, based on well-established understanding of scenarios. Since scenarios are often regarded as ``accessories'' or secondary artifacts, shared understanding of a scenario method has been lacking. Compared to other software and systems engineering artifacts, such as requirements/use-case specifications, test cases, and simulation input specifications, distinct characteristics of scenarios have not been adequately explained and provided. In other words, the term scenario has a variety of published definitions from a vast and bewildering array of domains~\cite{alexander2000scenarios_in_se, updegrove2018schema_based}.

Second, another major hurdle of applying the scenario method is the lack of a conceptual basis or framework for developing, selecting, and evaluating the methods. For an effective analysis of the method, theoretical or conceptual framework should be provided to enable the identification and organization of core concepts, which outlines the essential features of diverse methods. There have been various studies to define scenarios and build typologies in a number of engineering domains over the years~\cite{siegfried2012scenarios_in_military, siso2018guideline}. However, to the best of our knowledge, comprehensive investigation of scenario data and variables of scenario methods has not been conducted yet.

One of the best approaches to establish a shared understanding and a conceptual basis is to use metamodeling and conceptualization~\cite{durak2014scenariodevelopment,Karagiannis2013MetaModelingAA}, which define core concepts as meta-classes and analyze relationships between the classes. By providing a higher level/layer of abstract concepts and their associations, the conceptual basis is able to attain its extensibility and flexibility so that engineers can be systematically guided. To design a well-established framework, the concepts should be thoroughly identified, defined, and organized. Therefore, the conceptualization process should include the investigation of the concepts, relationships, and their attributes.

In this study, as a first step to establish a conceptual basis, a semi-systematic literature review on publications that study scenarios or utilize scenario methods is conducted. It allows us to grasp various meanings and purposes of employing scenario methods and to identify commonly (or frequently) used conceptual variables and data. Specifically, the data collected from the selected publications are defined as \textit{scenario variables} (SVs), and they are classified into 4 levels of classes (\textit{method-level}, \textit{suite-level}, \textit{scenario-level}, and \textit{event-level}) to define constructs of a scenario specification. Among 1071 publications searched, 354 publications were finally selected based on selection criteria. From the finally selected publications, 100 highly-related publications were inspected to collect the variables. 
This study further conceptualizes the SVs by meta-modeling to develop a \textit{Conceptual Scenario Model} (CSM), which can be used as a communication and analysis tool to establish and strengthen the shared understanding and to enable scenario assessment.

For the evaluation of our conceptual model and its constituent scenario variables, this study analyzes actual scenario instances developed in an automated driving system (ADS) domain. The ADS domain is an application domain that most actively employs scenario methods for various engineering tasks, such as safety analysis, simulation, and testing. In addition, there has been much effort to standardize scenario-based engineering, such as \textit{ASAM OpenSCENARIO} (and \textit{OpenDRIVE})~\cite{asam2020openscenario} and \textit{PEGASUS} method~\cite{mazzega2019pegasus}. Through the real-world (or simulation) scenarios for ADSs, applicability and validity of our conceptual model and variables are evaluated.

The remainder of this paper is structured as follows. Sections~\ref{sec_related_work} and \ref{sec_background} introduces related work and background knowledge, and Section~\ref{sec_overall_approach} presents an overall approach of this study. Section~\ref{sec_survey} conducts a semi-systematic literature review, and Section~\ref{sec_sv} defines scenario variables collected from the investigation. Section~\ref{sec_evaluation} performs a case study with real-world scenario methods/instances, and Section~\ref{sec_evaluation_discussion} analyzes and evaluates the applicability and expressiveness of the SVs. Section~\ref{sec_conclusion} concludes the paper and presents the future work.
\section{Related Work}
\label{sec_related_work}


This section introduces several scenario methods, including scenario-based/driven analysis, design, simulation, testing, and validation. To differentiate scenarios utilized in automated driving system (ADS) domains, the methods are classified into general-purpose (Section~\ref{sec_related_work_gp}) and ADS scenarios (Section~\ref{sec_related_work_ads}) as domain-specific scenarios.

\subsection{General-Purpose Scenario Development Methods}
\label{sec_related_work_gp}

General-purpose (GP) scenario methods typically aim to provide common standards and agreements for scenario specification and validation. Major advantages of utilizing GP methods are from (i) utilizing well-known modeling languages (e.g., graph-based, diagrammatic modeling languages) and (ii) intuitive and widely accepted semantics of the language.

The most traditional and simplest method to define scenarios might be using a graph or tree-based model to describe and test possible sequences of states or events/actions. By analyzing alternative paths (branches) of possible scenarios, functional behaviors of a system under study are defined as threads, and executable sequences in the exploration along the tree are defined as scenarios in \textit{Scenario Tree} (ST)~\cite{tsai2001scenariobased_regressiontesting}. By extending ST, \textit{Scenario Search Tree} (SST)~\cite{cunning2005automating_tg} was also developed to explicitly include conceptual variables apart from the functionality, states, or events. Although both approaches model system behaviors well (i.e., almost similar to a system's behavioral model), they have limitations in expressing contextual information and environmental conditions, which can influence the system behaviors.

For more sophisticated representation of scenarios, \textit{ACDATE/ Scenario model} specifies scenarios based on the \textit{Actors}, \textit{Conditions}, \textit{Data}, \textit{Actions}, \textit{Timing}, and \textit{Events} (ACDATEP if \textit{Policies} are included)~\cite{tsai2021scenario_based_mas}. This approach supports scenario-oriented requirements engineering and scenario planning through the \textit{Integrated ACDATE/Scenario Model} (IASM) of command \& control systems. By analyzing and formally specifying relationships between the system and the scenario, this model facilitates static analysis of the models to check the completeness \& consistency and analyze service properties, such as reliability. Although it is evident that this method has the expressiveness to support generic behavioral modeling of various system types, low-level (i.e., code-like) and fixed set of semantics limit flexible and extensible abstraction for the scenario specification.

To introduce and realize scenario-based programming, the visual languages \textit{message sequence charts} (MSCs) and \textit{live sequence charts} (LSCs) are used to describe flows of how the system (and its interface) has to react to user inputs~\cite{sikora2010msc, dan2012msc}. Like a UML sequence diagram, they contain participants (and their lifelines), environment, and interactions (messages) between them to depict runtime behaviors of a system. By adding liveness and execution semantics to behaviors of MSC, LSCs are able to graphically represent event patterns, conditions, constraints, and predicates (e.g., what is mandated or what is not allowed). 

Aside from these methods, many existing approaches have utilized variants of the UML/SysML-style sequence diagram (e.g., Action Sequence Charts (ASCs)~\cite{heymans1998ScenarioBasedTF}, Modal Sequence Diagram (MSD)~\cite{greenyer2017scenariotools, greenyer2015evaluating_formal}), semi-formal diagrams (e.g., process mining for scenario discovery~\cite{wang2014businessprocess}), and formal modeling languages (e.g., Petri Nets (PNs)~\cite{faria2014petrinets}, Hybrid Automata (HA)~\cite{alessandro2011hybrid_automata}, and Extended Finite State Machine (EFSM)~\cite{zhang2018correct_by_construction}). 


    
    
    
    

One limitation of GP methods is the high level of abstraction. Consequently, the semantics may not be specialized as domain-specific approaches (e.g., military~\cite{SISO2008MSDL, wittman2009msdl}, autonomous driving, aviation~\cite{jafer2016asdl, jafer2017asdl}, and programming/platform-specific scenarios (\textit{Gherkin Scenario}\footnote{\textit{Gherkin Syntax}, \url{https://cucumber.io/docs/gherkin/reference/}}, \textit{Scenario Modeling Language for Kotlin}~\cite{wiecher2020scil})) require. On the other hand, GP methods often focus on behavioral and interactive descriptions instead of provisioning contextual information. To judge from domain-specific perspectives, the missing or insufficiently identified contexts can lead to a decline in overall effectiveness of scenario methods.

\subsection{Scenario Methods in ADS Domains}
\label{sec_related_work_ads}

Domain-specific scenario methods have been frequently employed and utilized in domains that contain critical systems, such as safety and mission critical systems. In general, these critical systems are not developed by a single development team, but by numerous stakeholders and engineering groups from different backgrounds that participate in one large development project. Therefore, scenarios play a significant role in allowing engineers to clearly capture specifications and facilitate more visible communication between stakeholders, during the whole development phases. The ADS domain, where scenario methods are the most actively employed, also utilizes scenarios to engineer many critical features, dynamics and behaviors, processes, and regulations/policies, according to a given standard(s).

The \textit{ASAM OpenSCENARIO}~\cite{asam2020openscenario} is a standard by \textit{Association for Standardization of Automation and Measuring Systems} (ASAM) for scenario development of traffic simulation along with \textit{OpenDRIVE}\footnote{Open Dynamic Road Information for Vehicle Environment} and \textit{OpenCRG}\footnote{Open Curved Regular Grid}. Contrary to \textit{OpenDRIVE} and \textit{OpenCRG}, which provide static contents, \textit{OpenSCENARIO} provides vendor-independent dynamic traffic elements and maneuver libraries. Therefore, the family of standards supports multi-layered environment for scenario development and simulation to test, validate, and certify safety operations in driver assistance systems and autonomous driving environments. 
The standard mainly supports flexible modeling of automated driving maneuvers by providing storyboard-based logical scenarios. Consequently, constructs of this standard are very domain-specific (e.g., \textit{maneuver, trajectory, vehicle, driver, traffic environment}), which follow domain experts' knowledge and existing standards/regulations, rather than basing on generic concepts.

\textit{PEGASUS Method}~\cite{mazzega2019pegasus} is a method for generating logical scenarios and scenario-based testing of automated driving functions. Based on a given processing chain, the method defines three major components, which are \textit{events}, \textit{scenarios}, and \textit{test cases}. The process of PEGASUS method not only suggests the artifacts, but also includes concrete data (format), knowledge, models as databases. 
Also, \textit{6-Layer Model (6LM)}---from road network to digital information---was also introduced within the PEGASUS project to comprehensively engineer the ADS traffic environment for its operational safety.
Based on the PEGASUS method and 3 abstraction levels of scenarios defined by T. Menzel \etal~\cite{menzel2018scenarios}, a graphical modeling language/framework, called \textit{SceML}, was recently introduced~\cite{schutt2020sceml}. The SceML supports modeling of multiple abstraction levels and modularization of sub-scenarios using a graph-based scenario model.
    
Similarly, \textit{Scenario Description Language (SDL)} suggested by X. Zhang ~\cite{zhang2020sdl_for_driving} is a modeling language to specify scenarios of automated driving systems. The major strength of the SDL is the identification of various models and data (e.g., accident database, \textit{STPA}\footnote{System-Theoretic Process Analysis}) needed for the scenario development. The developed scenarios comprehensively include the scenery, dynamic elements, base scenarios/elements, and can define contextual and causal elements/factors accordingly. Their research focuses on the testing of control action specific to the driving system domain and possible unsafe causes/hazards (i.e., corner cases). Consequently, the scenario elements are defined specifically at a domain-level.

Compared to general-purpose methods, the scenario methods in ADS engineering fields have focused more on the contextual information, such as non-ego vehicles, traffic of the road network, and weather conditions. Also, similarly to other critical systems (e.g., military \& aviation fields), the ADS scenarios distinguish normal baseline scenarios from critical scenarios, according to the criticality (e.g., safety or mission) of behaviors under study. Even though the domain-specific approaches more practically support ontological analysis and application of domain knowledge, the lack of commonly shared conceptual variables leads to low extensibility and flexibility of the methods.
The introduced studies are just a few of scenario studies in ADS domains. Scene-focused scenarios (e.g., Ontology-based Scene Creation~\cite{bagschik2018ontology_based}, Scenic~\cite{fremont2019scenic}), infrastructure/environment-focused scenarios~\cite{wen2021virtual_scenario}, and risk/hazard-focused scenarios~\cite{bagschik2016hazardous_events} also hold important positions in scenario-based ADS testing and simulation. Some other scenario instances will be analyzed in Section~\ref{sec_evaluation}.


    
    
    
\section{Background}
\label{sec_background}


\subsection{Scenario and Scenario Method}
\label{sec_background_scenario}

As discussed in Section~\ref{sec_intro}, the terms \textit{scenario} and \textit{scenario method} have a variety of published definitions from a vast and bewildering array of domains~\cite{alexander2000scenarios_in_se, updegrove2018schema_based}. Since the term originated from film industry, a scenario originally refers to an outline for a screenplay, which is a script (prescribed form) of a series of scenes. The scenario is defined as ``\textit{a postulated sequence or development of events}'' in \textit{Oxford Dictionaries}, and the scenario typically refers to ``\textit{a description of how the view of the world changes with time, usually from a specific perspective},'' as the \textit{OpenSCENARIO} standard defines. On the other hand, some other publications simply define a scenario as ``\textit{evolution (temporal/causal development) of scenes}'' or ``\textit{a path of in a transition graph consisting of at least one state/event}.'' For better understanding of this paper and consistently using the terms, this section generally defines \textit{Scenario} and \textit{Scenario Method} in our way.

\subsubsection{Scenario}
\label{sec_background_scenario_scenario}

\begin{dfn}[Scenario]
    A \textit{scenario} (or a scenario specification) is a coherently described course of significant events to concretize paths of possible dynamics under a particular context(s), on the basis of specific purpose and hypothetical extrapolation.
\end{dfn}

The above definition includes four important keywords: course of events, possible dynamics, context, and hypothesis. The \textit{possible dynamics} of a system and its environment are represented by the \textit{course of events}, which describe an articulated behavior thread and its path. Here, the path can be probable, plausible, or just possible, depending on the level of complexity and uncertainty~\cite{keith2018scenario_development}. Also, a scenario should include specific \textit{contextual information}, which is expected to affect both the execution of a scenario and the occurrences of its events. By providing particular contexts to the dynamics description, a scenario plays a role as a device to capture and lucidly explain given specifications. Lastly, there is no scenario without a goal (i.e., \textit{hypothesis} made from a particular viewpoint); if there is no goal, it is just a plot/story. By determining a specific viewpoint and identifying solid objectives and their values, goal-oriented hypotheses can be defined. In other words, a hypothesis informs what an engineer want to observe/analyze/validate, thus they are frequently derived from system's goals, requirements, evidences (e.g., historical data), and theories. If a model is developed in accordance with the above requirements, it can be considered a scenario specification, but a set of specific semantics must be determined by an underlying formalism (i.e., a modeling language).

\begin{figure}
\centering
  \includegraphics[width=\columnwidth]{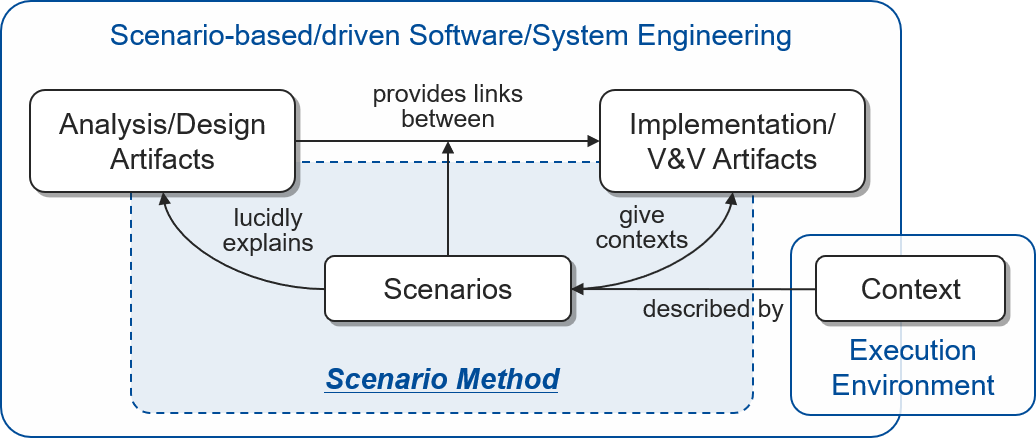}
  \caption{Illustration of a scenario method}~\label{fig_scenario_method_illustration}
\end{figure}

At this level/stage, types, uses, and purposes of scenarios are not determined. Although there have been many attempts to define the typologies of scenarios~\cite{vannotten2003updated_typology, vannotten2006typology_approaches, marco2019scenario_planning}, two classification methods are primarily used.
The first approach, mainly used in military and aviation domains, classifies scenarios as \textit{operational}, \textit{conceptual}, and \textit{executable} scenarios~\cite{siegfried2012scenarios_in_military, msg086_2014_guideline, durak2014scenariodevelopment, Jafer2018SchemabasedOR}, based on the maturity level of the scenarios. The other approach, mainly used in automated driving system (ADS) fields, distinguishes three types of scenarios as \textit{functional}, \textit{logical}, and \textit{concrete} scenarios~\cite{menzel2018scenarios, menzel2019functional_logical}. 
Both approaches divide the levels in terms of the abstraction levels and explanation methods of scenarios. For example, \textit{operational} and \textit{functional} scenarios are typically written as a narrative and generally explained by domain experts. On the other hand, \textit{executable} and \textit{concrete} scenarios represent low-level data and execution mechanism (e.g., algorithm), and they explain how an execution method (e.g., real-world, simulator, testing engine, training) should run the scenario. It is clear that the specification and modeling of multiple abstraction levels of scenarios should be appropriately supported for more systematic scenario-based engineering.




\subsubsection{Scenario Method}
\label{sec_background_scenario_scenario_method}

\begin{dfn}[Scenario Method]
    A \textit{scenario method} is any engineering method or approach that develops, uses/utilizes, manages a scenario(s) for a particular engineering purpose. 
\end{dfn}

Scenarios of a scenario method are typically products of concretization. They provide links between analysis/design artifacts and executable artifacts (e.g., implementation or V\&V artifacts) by giving logical or concrete contexts to them. As Figure~\ref{fig_scenario_method_illustration} shows, a scenario plays a role as a proxy to capture the specification (e.g., analysis \& design artifacts) and give contexts to executable models. The contextual information is derived from the analysis of available execution environment, such as testing, simulation, real-world execution, or training. By delivering logical or concrete contexts, scenarios provide links and narrow the gap between engineering activities of different phases. Also, a \textit{scenario engineer} refers to an engineer who utilizes (or is related to) the scenario method, and a \textit{scenario stakeholder}, including scenario engineers, refers to a person who can either affect or be affected by the method.


\begin{figure}
\centering
  \includegraphics[width=0.7\columnwidth]{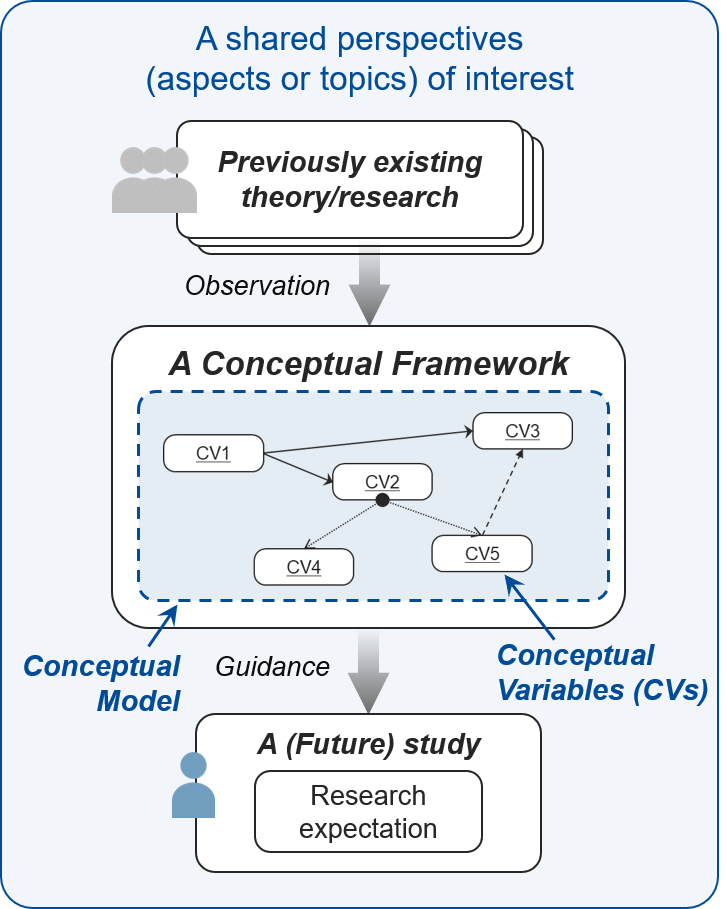}
  \caption{High-level description of the conceptual framework for scenario methods}~\label{fig_csf_roles}
\end{figure}

\begin{figure*}
\centering
  \includegraphics[width=\textwidth]{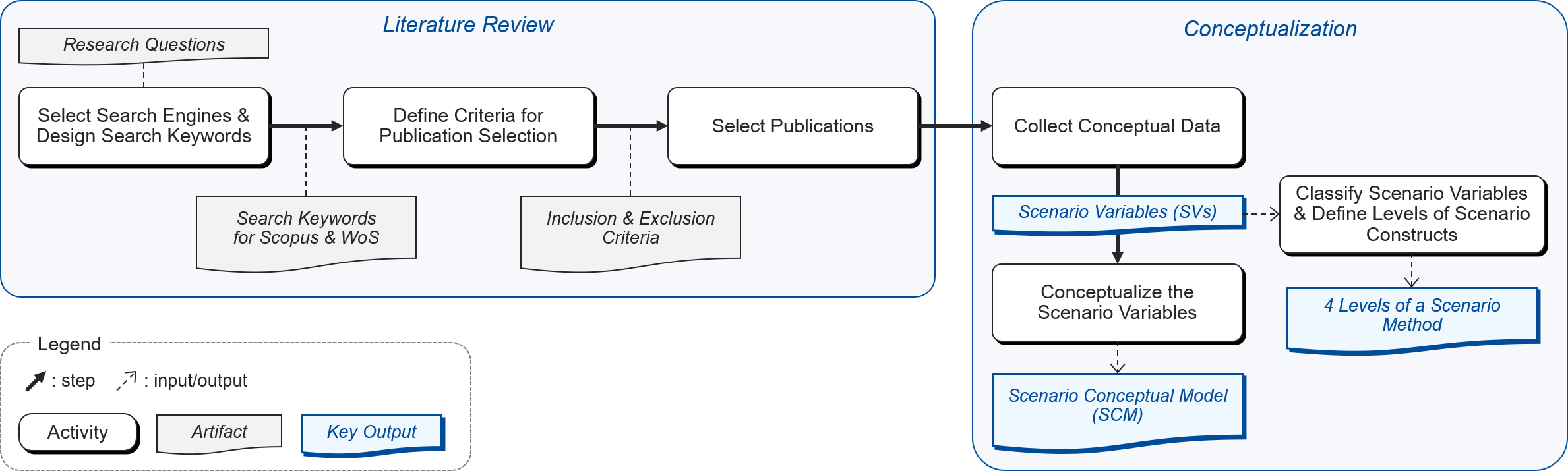}
  \caption{Overall approach of this study}~\label{fig_overall_approach}
\end{figure*}

\subsection{Conceptual Framework}
\label{sec_background_cf}


A conceptual framework provides a ground to explore diverse perspectives on a subject research area of interest by explaining the importance of a topic in both practical and theoretical sense. To provide a comprehensive understanding of the universe of discourse (e.g., phenomenon), the conceptual framework provides an interpretative approach by articulating concepts as constructs in which each concept plays an integral role ~\cite{jabareen2009conceptualframework}. Major roles are (a) to isolate \textit{key variables} (i.e., core concepts) as a focal framework and (b) conceptualize the variables so as to focus and set boundaries. As Figure~\ref{fig_csf_roles} illustrates, intellectual traditions can be empirically studied (and also observed) to establish a conceptual framework that organizes the things of the past, such as terms/concepts, data, models, experience, and experiments, etc. By detailing methods to answer research questions of the research area, conceptual frameworks have been suggested to find solutions specific to a set of problems or ideas from a practical perspective.

Our ultimate goal is the construction of a conceptual framework for scenario methods. Consequently, key variables of the scenario methods (i.e., \textit{things of the past} mentioned above) first needed to be systematically collected and conceptualized. The subject research area that this study focuses on is ``scenario methods." The particular target aspect is the development of scenarios, such as modeling and specification method. By distinctly setting a conceptual boundary for scenario methods, future studies can be guided using our conceptual scenario framework. In order to ensure reflexivity and dialogic engagement, selected variables must be able to cultivate research questions of scenario methods and play a role as a compass system to establish practical and extensible knowledge.


\section{Overall Approach}
\label{sec_overall_approach}

In an effort to establish a conceptual basis for scenario methods, an in-depth and rigorous investigation of various types of scenario methods needs to be preceded. To extensively investigate the methods and set clearer boundary of the universe of discourse, this study conducts a semi-systematic literature review to collect conceptual variables and answer research questions. Through the survey, meaningful issues, challenges, and future research directions are also derived and discussed.

This study follows the three steps in Figure~\ref{fig_overall_approach}. \textit{Step 1. Collection of publications}: We design search queries and selection criteria (i.e., inclusion and exclusion criteria) to systematically obtain as diverse scenario methods and concepts as possible. \textit{Step 2. Identification of conceptual variables from the publications}: Our literature review aims to systematically collect information and data for the development of a conceptual framework consisting of conceptual variables. Therefore, the survey is conducted based on specific research questions and 4 levels of properties, to generate meaningful analysis and statistics. \textit{Step 3. Conceptualization of the variables as a conceptual model}: By analyzing the scenario variables collected at the previous step, this study defines a semantic domain, which a scenario method must provide, by building a \textit{Scenario Conceptual Model} (SCM). Scenario engineers can get help in applying scenario methods (and developing scenarios) through the variables of the SCM in the future.
\section{Literature Review of Scenario Methods}
\label{sec_survey}

\subsection{Design of Literature Review}
\label{sec_survey_design}

Targets of this survey are (a) scenarios defined or specified/modeled in selected publications, and (b) scenario-based/driven engineering approaches (i.e., methods, techniques, methodologies) suggested or utilized in the reviewed publications. The ultimate goal of this literature review is to develop a conceptual basis for scenario methods. Therefore, this study focuses on identifying as much conceptual data and defining them as conceptual \textit{scenario variables}, shortly SVs.

\subsubsection{Research Questions}
\label{sec_survey_design_questions}

This study mainly focuses on semantics to be considered in scenario methods, and thus diverse properties of scenarios defined in the publications are collected and analyzed. To recognize them, our review will answer the following key questions: 
\begin{enumerate}[\bf RQ1:]
\item How do the publications define scenarios and what data and properties are related to scenarios?
\item How can the investigated conceptual variables be classified into different levels of constructs (i.e., concepts, properties, or data values) and be conceptualized?
\item What are the remaining challenges to design or apply a scenario method?
\end{enumerate}

To answer the first question, all relevant scenario data are collected from finally selected publications. They are defined as \textit{scenario variables (SVs)} and possible values that can be assigned to the variables (See Section~\ref{sec_sv}). The second question is answered by classifying and conceptualizing the variables, and multiple levels and a conceptual model of a scenario method are developed to answer the question in Section~\ref{sec_sv}. In Section~\ref{sec_evaluation_discussion_challenges}, the third question will be answered by discussing salient issues and challenges engineers may encounter during the conceptualization.

\begin{figure}
\centering
  \includegraphics[width=\columnwidth]{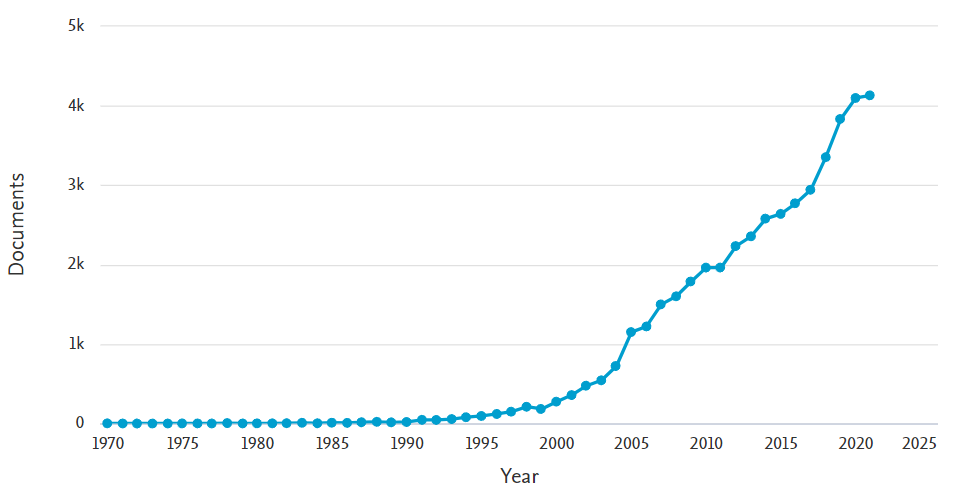}
  \caption{Increasing trend of scenario methods in software and systems engineering fields (This chart is made by \textit{Scopus})}~\label{fig_documents_by_year}
\end{figure}

\subsubsection{Search Engines and Search Keywords}
\label{sec_survey_design_search}

In the survey, initial raw publications were first collected using the most well-known search engines in the engineering and science fields, \textit{Scopus} and \textit{Web of Science} (WoS). Because \textit{Google Scholar} only supports simple keywords and filtering options, it provides relatively less systematic search results. Therefore, we used \textit{Google Scholar} only as a supplementary engine for tracking additional references from the selected publications.

The first step of our literature review process is the construction of search queries and keywords. To design finer queries and to preliminarily check how often scenarios are used in software/systems engineering domains, we searched arbitrary publications only using the keywords of ``\textsf{software engineering},'' ``\textsf{system(s) engineering},'' and ``\textsf{scenario}'' on both search engines\footnote{This search was done on 4th February, 2022.}. The initial search returned 46,842 results (46,018 from \textit{Scopus} and 824 results from \textit{WoS}) when ``\textsf{software engineering}'' and ``\textsf{scenario}'' were included in the keyword, and 42,850 results (42,060 from \textit{Scopus} and 790 results from \textit{WoS}) are returned when ``\textsf{system(s) engineering}'' and ``\textsf{scenario}'' were included. The results showed that there is a growing evidence that both engineering domains have increasingly utilized scenario methods over the years (See Figure~\ref{fig_documents_by_year}).

Because the initial search did not set the boundaries of engineering activities and assumptions, a substantial amount of results were returned accordingly. To research a tuned set of scenarios and scenario methods used in software/systems engineering domains, we refined the search and determined three main keywords. First, like the initial search, ``\textsf{software/system(s) engineering}'' was included in the search term to limit the engineering domain of publications to software and systems engineering studies. Second, ``\textsf{scenario}'' was certainly included to research scenarios and scenario-based methods, techniques, and methodologies. To get more elaborate results, ``\textsf{scenario-based/driven}'' and ``\textsf{event}'' were also included in the actual search query to retrieve scenario methods that consider events. Finally, ``\textsf{requirement},'' ``\textsf{validation},'' ``\textsf{test},'' and ``\textsf{simulation}'' were included in the query, because they were deemed the four most representative engineering activities employing scenario methods. 

Following are the search queries for \textit{Scopus} and \textit{Web of Science}, respectively\footnote{Note that the research area was limited to \textit{computer science} and \textit{engineering} (i.e., \textit{COMP}) for both search engines.}.

\begin{spacing}{0.8}
\begin{quote}
{\centering\small\textsf{ALL("software engineering" OR "system engineering" OR "systems engineering")
AND TITLE-ABS-KEY(("scenario*" OR "scenario-based" OR "scenario-driven") AND "event*")
AND TITLE-ABS-KEY("requirement*" OR "validation*" OR "test*" OR "simulation*")
AND PUBYEAR $>$ 1999
AND (LIMIT-TO (SUBJAREA , "COMP")) 
AND (LIMIT-TO (LANGUAGE , "English"))}}
\end{quote}

\begin{quote}
{\centering\small\textsf{"software engineering" OR "system engineering" OR "systems engineering" (All Fields) 
and "scenario*" OR "scenario-based" OR "scenario-driven" (All Fields) 
and "event*" (All Fields) 
and "requirement*" OR "validation*" OR "test*" OR "simulation*" (All Fields) 
and 2000-2021 (Year Published) 
and English (Language)
and Engineering or Computer Science (Research Areas)}}
\end{quote}
\end{spacing}

The reason why this is a \textit{semi-systematic literature review} is that (a) we only use two major search engines (\textit{Scopos}, \textit{WoS)} and (b) the search keywords are not incrementally designed or refined during the survey process. In other words, our survey does not use a snowball method to find more literature from selected publications. Therefore, the selection process, introduced in Section~\ref{sec_survey_selection}, only excludes publications that are away from our interests and intentions, which are aligned with the questions in Section~\ref{sec_survey_design_questions}.

\subsubsection{Selection Criteria}
\label{sec_survey_design_selection_criteria}

Based on the target questions in Section~\ref{sec_survey_design_questions}, inclusion and exclusion criteria are defined to efficiently examine the publications.

\paragraph{Inclusion Criteria} This review aims to include the following publications to collect data, concepts, and values of scenarios or scenario methods.

\begin{new_itemize}
    \item Studies of scenario development, such as modeling and specification methods (and languages), techniques, processes, and methodologies
    \item Scenario-based or -driven engineering studies, which explicitly specify scenarios for specific engineering purposes, such as scenario-based requirements engineering, validation, design, testing, simulation, and verification
\end{new_itemize}

\paragraph{Exclusion Criteria} This survey excluded the publications collected from the search engines based on the following criteria. There are two levels of exclusion processes, which are \textit{title-abstract-exclusion-criteria} for rounds 2 and 3, and \textit{ARP-FSP-exclusion-criteria} for rounds 4 and later (See Section~\ref{sec_survey_selection}).

\begin{new_itemize}
    \item \textit{title-abstract-exclusion-criteria}: Exclusion criteria for title and abstract review
    \begin{new_itemize}
        \item Inapt publication type (e.g., whole proceedings, an entire book, newspaper articles, web pages, etc.)
        \item Unrelated engineering domain, which is not related to software or systems engineering (e.g., chemical, biological, medical engineering or non-engineering publications)
        \item Unrelated system or application domain (e.g., political or organizational system, international ecosystem)
        \item Unrelated approach, which does not utilize scenarios for an engineering purpose, and less than or equal to 3 pages
    \end{new_itemize}
    
    \item \textit{ARP-FSP-exclusion-criteria}: Exclusion criteria for reviewing abstract-reviewed publications (ARPs) and finally-selected publications (FSPs)
    \begin{new_itemize}
        \item Scenarios only used as a term to simply represent a system, a system type, a paradigm, or a case
        \item Scenarios (or scenario methods) not explicitly used or mentioned (i.e., unable to retrieve in a document)
        \item Absence of scenario instances or insufficient semantic data of scenarios or events
        \item Totally informal scenarios (i.e., narrative descriptions)
    \end{new_itemize}
    
    \item \textit{relevance-criteria}: Criteria to pick out more relevant publications from the FSPs
    \begin{new_itemize}
        \item The number of search result of keywords (i.e., appearance count in each publication): \textit{numOfScenario}, \textit{numOfEvent}
        \item Presence of a scenario-describing figure(s) or table(s)
    \end{new_itemize}
\end{new_itemize}

\begin{figure}
\centering
  \includegraphics[width=\columnwidth]{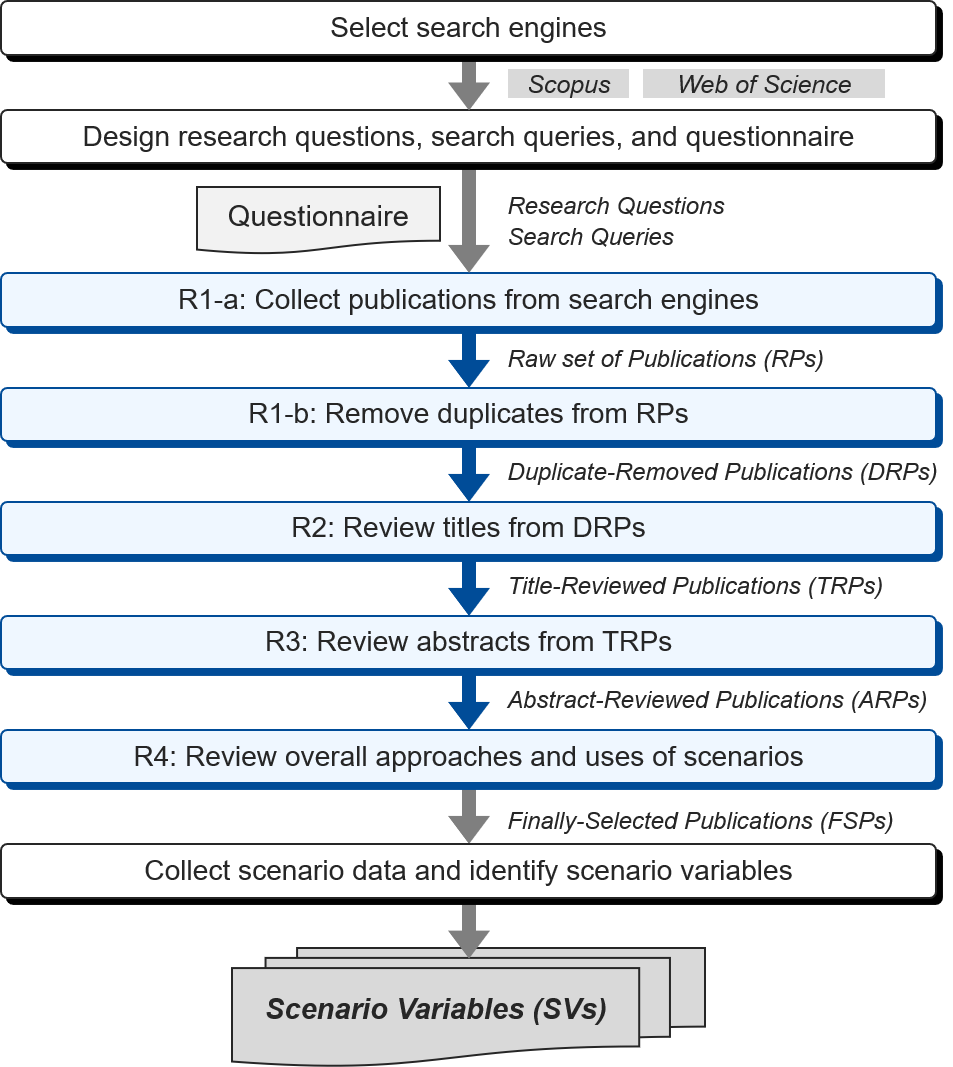}
  \caption{A publication selection process for the literature review}~\label{fig_survey_process}
\end{figure}

\subsection{Selection of Publications \& Data Collection}
\label{sec_survey_selection}


This section describes the overall process of collecting and reviewing publications for the survey, which is depicted in Figure~\ref{fig_survey_process}. 

\paragraph{Round 1 (R1): Collect Publications and Remove Duplicates} We initially collected publications from the search engines using the search terms. As explained in Section~\ref{sec_survey_design_search}, \textit{Scopus} and \textit{WoS} were selected as search engines, and the search queries are designed for both engines to collect relevant references. The initially collected publications are called \textit{raw publications (RPs)} and the set of publications after the duplicate removal is called \textit{duplicate-removed publications (DRPs)}.

\paragraph{Round 2 (R2): Review Titles} In the following steps, we reviewed and excluded publications based on the selection criteria for each step. In R2, we reviewed the title of the DRPs based on the \textit{title-abstract-exclusion-criteria} to obtain \textit{title-reviewed publications (TRPs)}. Major purposes of reviewing titles are as follows. First, this selection was intended to remove improper publication types. When searching only with keywords, the entire proceedings of a conference or an entire book may be included, which should be excluded. Second, publications that study either unrelated engineering domains or unrelated system/application domains should be removed from the publication list. Since they are more likely to be out of scope or not fit the purpose of this investigation, they are simply considered as unrelated studies. However, because the titles alone do not provide detailed information on the purposes and application domains, we additionally excluded the publications by conducting the abstract review in R3.

\paragraph{Round 3 (R3): Review Abstracts} In R3, we reviewed abstracts of TRPs, along with their introduction and conclusion sections, if needed. Like R2, this round also uses the \textit{title-abstract-exclusion-criteria} to exclude less relevant publications and generate a set of \textit{abstract-reviewed publications (ARPs)}. Because we could deduce details about the complex information and reasons for utilizing scenarios from the abstract sections rather than from only the titles, we manually reviewed abstract sections of TRPs.

\paragraph{Round 4 (R4): Review Overall Approaches and Uses of Scenarios} In this final selection round, key parts of the ARPs were reviewed based on the \textit{ARP-FSP-exclusion-criteria}. While reviewing overall approaches, we manually searched ``scenario" and assessed whether scenarios and scenario instances are explicitly defined and whether they are developed and prepared to serve the authors' purpose (i.e., course of events to describe dynamics and contexts). Even though most of the previously reviewed publications contained the term ``scenario," some of them did not use scenarios explicitly or scenarios sometimes did not provide enough semantic information (e.g., flow, contexts, events/actions, conditions, etc.). Since it is needed to verify the partial information that can be obtained from the abstracts and overall approaches of the publications, we had to confirm specific uses of scenarios. Through this final round, this survey obtained a set of \textit{finally-selected publications (FSPs)}, which are reviewed by full read. After collecting the FSPs, for a feasible review, 100 most-relevant publications are prioritized for the full-read review, and they are listed in Appendix~\ref{appendix_a} (For convenience, we call \textit{FSP-100} simply as \textit{FSP}).

Also, the overall description and the exact numbers of selected and excluded publications are summarized in Table~\ref{tab_selection_rounds}. The overall process for data collection is illustrated in Figure~\ref{fig_survey_process}. First, any type of ``conceptualizable'' data constituting or surrounding scenarios (and scenario methods) were identified, which is called \textit{raw data}. After the initial collection, the raw data can be identified as either \textit{variables} or \textit{possible values} of the variables.

\begin{table}
\caption{Selection of publications from Round 1 to Round 4}
\vspace{-5pt}
\centering
\resizebox{1.05\linewidth}{!}{
\begin{tabular}{lllll} 
\toprule
\textbf{Round}                                                                                 & \textbf{Criteria}                                                                                              & \textbf{Output}                                                                                  & \textbf{Selected} & \textbf{Excluded}  \\ 
\midrule
\begin{tabular}[c]{@{}l@{}}R1-a: Collect\\Publications\end{tabular}                            & \textit{-}                                                                                                     & \begin{tabular}[c]{@{}l@{}}\textit{Raw Publications}\\\textit{(RPs)}\end{tabular}                & 1071              &                    \\ 
\midrule
\begin{tabular}[c]{@{}l@{}}R1-b: Remove\\Publications\end{tabular}                             & \multirow{2}{*}{\begin{tabular}[c]{@{}l@{}}\textit{title-abstract-}\\\textit{selection-criteria}\end{tabular}} & \begin{tabular}[c]{@{}l@{}}\textit{Duplicate-removed}\\\textit{Publications (DRPs)}\end{tabular} & 992               & 79                 \\ 
\cmidrule{1-1}\cmidrule{3-5}
R2: Review Titles                                                                              &                                                                                                                & \begin{tabular}[c]{@{}l@{}}\textit{Title-reviewed}\\\textit{Publications (TRPs)}\end{tabular}    & 851               & 141                \\ 
\midrule
R3: Review Abstracts                                                                           & \multirow{2}{*}{\begin{tabular}[c]{@{}l@{}}\textit{ARP-FSP-}\\\textit{selection-criteria}\end{tabular}}        & \begin{tabular}[c]{@{}l@{}}\textit{Abstract-reviewed}\\\textit{Publications (ARPs)}\end{tabular} & 765               & 86                 \\ 
\cmidrule{1-1}\cmidrule{3-5}
\begin{tabular}[c]{@{}l@{}}R4-a: Review\\Overall Approaches\\\& Uses of Scenarios\end{tabular} &                                                                                                                & \begin{tabular}[c]{@{}l@{}}\textit{Finally Selected}\\\textit{Publications (FSPs)}\end{tabular}  & 354               & 411                \\ 
\midrule
\begin{tabular}[c]{@{}l@{}}R4-b: Select 100\\Most-Relevant \\Publications\end{tabular}         & \textit{relevance-criteria}                                                                                    & \textit{FSP-100}                                                                                 & 100               & 254                \\
\bottomrule
\end{tabular}
\vspace{-8pt}
\label{tab_selection_rounds}
}
\end{table}


\section{Conceptualization of Scenario Variables}
\label{sec_sv}

To identify and collect core concepts related to general scenario methods, literature review was conducted, and raw data was collected from the selected publications in the previous section. On the basis of the collection, this section conceptualizes the data as well-established variables and classifies them into different levels.

\subsection{Abstraction of Scenario Data as Conceptual Scenario Variables (SVs)}
\label{sec_sv_abstraction}

A \textit{Scenario Variable} (\textit{SV} for short) is any concept related to scenarios or scenario methods that can have a concrete value (or a set of values). While collecting the SVs, there were many relations of inclusion between the SVs in terms of their semantics. For example, \textit{EventTemporal(Attribute)} can include \textit{EventTemporalData}, which can have a set of concrete data variables, such as \textit{start\_time}, \textit{duration}, or \textit{delay}. In this case, the SVs are further classified into \textit{primary variables} and \textit{subordinate variables}, as Table~\ref{tab_scenario_Variables} summarizes. The data collection and variable identification processes were manually conducted by the authors, and the detailed definitions for each variable type are introduced after the classification of the SVs (See Section~\ref{sec_sv_levels}). 

\begin{table*}
\caption{Scenario variables (SVs) identified and collected from the literature review}
\centering
\resizebox{\linewidth}{!}{
\begin{threeparttable}{
\begin{tabular}{lll} 
\toprule
\textit{\textbf{Level*}} & \textit{\textbf{Primary SV}}                                    & \textit{\textbf{Subordinate SVs}}                                                                                                                                                                                                                                                                                                                                                                                                                                                                                                                                             \\ 
\midrule
\multirow{3}{*}{M}      & \vcell{MethodPurpose}                                          & \vcell{\textcolor[rgb]{0.149,0.149,0.149}{TargetProblem}\textcolor[rgb]{0.149,0.149,0.149}{,~}\textcolor[rgb]{0.149,0.149,0.149}{TargetVisionIntent}\textcolor[rgb]{0.149,0.149,0.149}{~(}\textcolor[rgb]{0.149,0.149,0.149}{UseOfScenario}\textcolor[rgb]{0.149,0.149,0.149}{),~}\textcolor[rgb]{0.149,0.149,0.149}{TargetWoI}\textcolor[rgb]{0.149,0.149,0.149}{, Hypothesis,~}\textcolor[rgb]{0.149,0.149,0.149}{StrategyTactic}\textcolor[rgb]{0.149,0.149,0.149}{, Assumption}}                                                                                  \\[-\rowheight]
                        & \printcellmiddle                                                & \printcellmiddle                                                                                                                                                                                                                                                                                                                                                                                                                                                                                                                                                              \\
\bottomrule
\end{tabular}

\begin{tablenotes}
      \small
      \item *M(-Lv): Method-level, St(-Lv): Suite-level, Scn(-Lv): Scenario-level, Evn(-Lv): Event-level
\end{tablenotes}
}
\end{threeparttable}
}
\label{tab_scenario_Variables}
\end{table*}

\begin{figure}
\centering
  \includegraphics[width=\columnwidth]{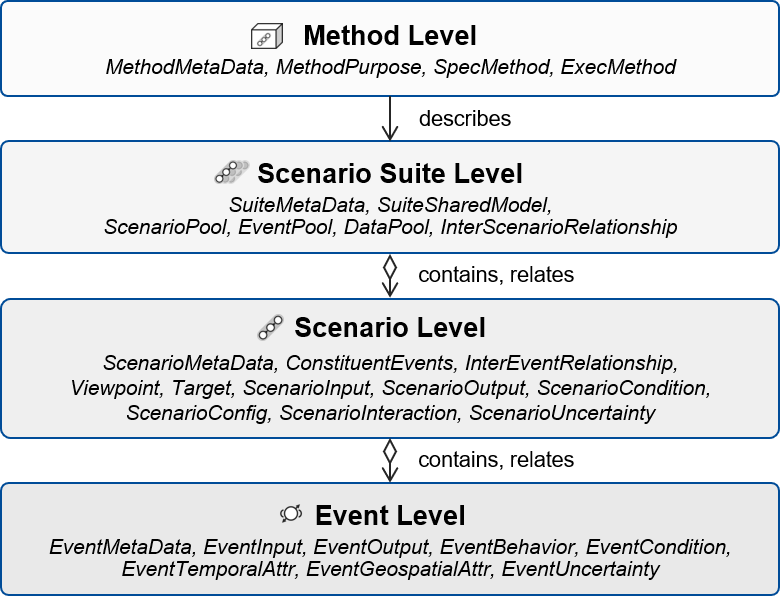}
  \caption{Four levels of scenario constructs}~\label{fig_four_levels}
\end{figure}

\subsection{Classification of SVs to Four Levels}
\label{sec_sv_levels}

The SVs collected from the FSPs were categorized according to four levels of a scenario method. 
Figure~\ref{fig_four_levels} portrays the hierarchical structure of the levels and their key elements. The biggest reason for classification is to distinguish the variables in terms of their shared aspects and to give communal rationale for each level. In other words, each level is expected to be engineered by different types of stakeholders to deal with more focused issues, knowledge, and information (e.g., a system engineer has much knowledge about system states and features (actions or functions) to specify the \textit{event-level} scenario constructs). Following are the descriptions of the levels and SVs included in each level. The primary SVs are defined in Appendix~\ref{appendix_b}.

\subsubsection{Method-level SVs}
\label{sec_sv_levels_method}

The highest-level of SVs are related to the development or selection of a scenario method to be employed for an engineering purpose(s). Specifically, method-level SVs elicit dialogic engagement of different stakeholders to determine overall goal and scope of a scenario method to be utilized. An appropriate scenario method is determined to analyze a universe of discourse, which consists of relevant goals/requirements (e.g., safety requirements, properties, or standards), plans, required decisions, technologies, and potential risks/hazards at a project level.

To provide overall information and understanding of a scenario method, method-level variables are mainly involved in the meta-data of a scenario method (\textit{MethodMetaData}), a purpose(s) of the method (\textit{MethodPurpose}), a specification method (\textit{SpecMethod}), and an execution method (\textit{ExecMethod}).

\subsubsection{Scenario Suite-level SVs}
\label{sec_sv_levels_suite}

Multiple scenarios of a scenario suite (i.e., a set of coherent scenarios) share the same viewpoint and its goals/values. Therefore, scenario engineers need to systematically identify and analyze the information shared among multiple scenarios to make a coherent and aligned scenario set. For this reason, the most important activity while collecting the suite-level variables is to reasonably determine a target(s) and an objective(s) by narrowing the gap between the scenario engineers from different backgrounds. After setting the target, the scenario development should be followed by the ontological analysis (e.g., system ontology, environment ontology, infrastructure ontology) and relevant data investigation to clearly determine a boundary the scenario method deals with.

For the acquisition of suite-level information, shared models (including knowledge and data) and potentially available components, such as reusable scenarios, events, and data (\textit{ScenarioPool}, \textit{EventPool}, and \textit{DataPool}, respectively) are identified. The most representative suite-level models are goal models, ontologies, and shared inputs/knowledge/data. The selection of the suite-level information is determined by the selected scenario method. Another important information at the suite level is \textit{InterScenarioRelationship}, which relates two or more scenarios to be incorporated as a single suite. The relationship spans from simple associations and dependencies to causal, concurrent, alternative, and interaction relationships.

\subsubsection{Scenario-level SVs}
\label{sec_sv_levels_scenario}

On the basis of information acquired as the higher level variables, scenario-level SVs focus on a single scenario specification. According to the specification method mentioned in Section~\ref{sec_sv_levels_method}, an abstraction level and semantics (i.e., contents of scenarios) can be defined differently. Also, some contents need to be included and specified in a way that an execution method (e.g., testing or simulation engine) requires.

Major scenario-level variables include the meta-data of an individual scenario (\textit{ScenarioMetaData}), its constituent events (\textit{ConstituentEvents)}, relationships among the events (\textit{InterEventRelationship}), a target(s) to be modeled (\textit{ScenarioTarget}), inputs and outputs (\textit{ScenarioInput}, \textit{ScenarioOutput}), scenario-level pre-/post-conditions (\textit{ScenarioCondition}), scenario-level configurable data (\textit{ScenarioConfig}), uncertainties to be considered (\textit{ScenarioUncertainty}), and so on.

\subsubsection{Event-level SVs}
\label{sec_sv_levels_event}

The lowest and the most detailed level of information is described by a set of event-level SVs, which represent behavioral aspects and dynamics of a scenario. As conventional modeling approaches define, an event refers to any occurrence that a system is designed to respond to and is denoted by its name and specific action(s). An event can be executed by either an internal trigger or an external trigger only if its precondition is met. Several studies distinguish event, act, action, activity, and stimuli, with respect to their different causes and types. Also, some studies define an action as an internally-triggered active behavior and an event as an external factor interacting with a system. Although we agree with the need of diverse definitions of behavior occurrences/executions, our conceptual framework defines the variables based on an assumption that all the behavioral information and data is an event and its occurrence. Also, concrete scenarios include concrete data values for the event occurrences or action executions at a lower level than the event-level variables. However, our conceptualization abstracts the data-level variables as the parameters or inputs of the events. Event-level variables represent singular behavioral aspects regardless of the types of events, such as \textit{act}, \textit{action}, \textit{stimuli} and \textit{internal/external change of states}.

\begin{figure*}
\centering
  \includegraphics[width=\textwidth]{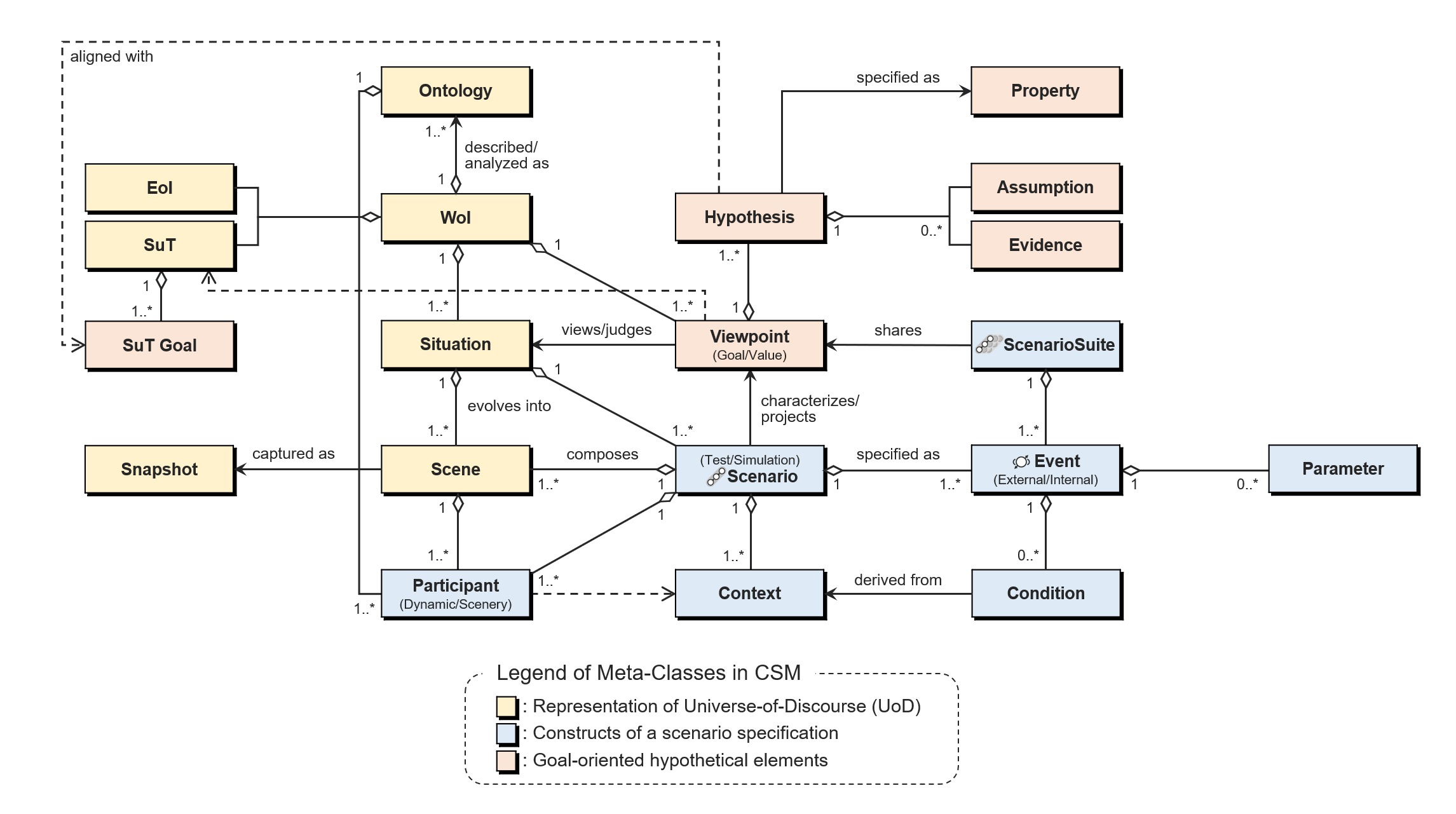}
  \caption{Conceptual Scenario Model (CSM)}~\label{fig_csm}
\end{figure*}

\subsection{Conceptual Scenario Model (CSM)}
\label{sec_sv_conceptual_model}

On the basis of collected variables, we conceptualize and elaborate them as a \textit{Conceptual Scenario Model (CSM)}, as the meta-model of Figure~\ref{fig_csm}\footnote{To provide a compact meta-model, attributes of the CSM meta-classes are omitted from the CSM diagram.}.
By defining the formerly defined SVs as \textit{meta-classes}, the CSM is designed to conceptualize crucial information of a scenario method and specification. The CSM has three types of meta-classes, which are described by different colors of boxes in the figure. Also, CSM aims to systematically define relationships between the meta-classes based on the four-levels defined for the SVs to be utilized as constructs for the scenario development. Although a specific guideline to specify a real scenario is not provided by simply identifying the concepts, this paper aims to provide a basis for developing a scenario specification method, specialized for a specific engineering purpose or application domain by defining the CSM.

According to the scenario definition in Section~\ref{sec_background_scenario_scenario}, a meta-model for specifying scenarios should include four key meta-classes to satisfy scenario requirements: \textit{Goal/Hypothesis}, \textit{Path/Flow} (i.e., course of events, possibilities), \textit{Context}, and \textit{Constituent Events}. In addition, according to the range and type of information focused in the scenario levels, it contains the following meta-classes: \textit{World-of-Interest (WoI)}, \textit{Situation}, \textit{Scene}, and \textit{Dynamics}. Therefore, the most important role of CSM is to relate and organize these meta-classes extracted from two sources to complement each other.

First, conceptual analysis of the WoI should be preceded in applying a scenario method. The most representative method to determine the boundary of WoI is ontology, which is an explicit specification of a conceptualization~\cite{gruber1995ontology}. The ontology is utilized as a framework that unifies different points of view on an area of interest without specifying a particular perspective. By systematically defining ontologies, the system and system environment can be characterized. An ontology model not only conceptualizes and assembles entities, but also establishes independent information management system, provides consistency of information utilized in various engineering activities and phases, and improves interoperability. Therefore, when systematic information of the WoI's operational domain (e.g., taxonomy) is provided, more complete and reasonable construction of ontology is possible. For example, for WoI analysis of Highway Autonomous Driving, ontologies such as \textit{highway ontology} (e.g., road network, traffic, regulatory elements), \textit{vehicle ontology} (e.g., driving functions, maneuvers), and \textit{weather ontology} (e.g., temperature, humidity, precipitation) can be constructed. In addition, ontological analysis for vehicle safety can be guided through \textit{Operational Design Domain} (ODD)\footnote{SAE J3016: Taxonomy and Definitions for Terms Related to Driving Automation Systems for On-Road Motor Vehicles (Revised on 2021-04-30)}. However, from a scenario engineer's perspective, the WoI components are often black-box, which may hinder the analysis of internal properties of entities solely based on ontology.

Second, main contents constituting scenarios and purposes of scenario specification are extracted from the specific goals and hypotheses of a scenario method, which are aligned with the goals of a System-of-Interest (SoI) under study (simply a \textit{System under Test\footnote{Here, we consider the term `test' as any type of analysis activity, such as simulation, testing, and verification.}} (SuT)). Consequently, many conventional SE processes guide use case and scenario specification based on the SoI goals or requirements specification. Each hypothesis is defined differently depending on the viewpoints despite the same WoI. Based on the viewpoints, key meta-classes of specific path/flow, context, and event can be defined (or articulated). A WoI contains various situations, thus setting the viewpoint allows analysis and specification of the target situation and behaviors of the SOI (SUT).

Once the viewpoint is configured for the WoI, the dynamics to be specified, analyzed, and observed through scenarios need to be defined. Usually defined as path/flow, dynamics are described as a sequence of events or sequence of scenes, and their granularity often aligns with a use case, a task, or a mission. Scenario-level context is defined to incorporate situations interacted or affected by the execution of scenarios. For describing a particular context, methods such as context diagram and (runtime) environment analysis can be utilized, which aims to analyze conditions, constraints, and assumptions affecting the flow and execution of behaviors from a specific viewpoint. Even when specifying identical system behaviors, alternative paths and hypotheses can be defined and validated within a scenario suite based on various contexts.

The last level of scenario specification is event, which specifically defines behavioral information and properties. Event-level primary SVs collected from the literature review are input, output, behavior, interaction, temporal information, geospatial information, and uncertainty. These SVs are utilized as data for real execution of events and configured as a parameter, which is one of the event-level inputs. Event-level context contains trigger conditions for a precondition and event transition for internal/external events to occur, which are defined as the \textit{EventCondition} variable (See Table~\ref{tab_scenario_Variables}). For example, an ADS scenario specifies events such as driving functions, maneuvers (e.g., full-brake), external events (e.g., pedestrian crossing), and environmental event (e.g., weather change). Each event can contain both temporal and spatial information and requires logical/concrete parameters (e.g., vehicle performance, acceleration rate, initial traffic scene) and contextual information for realistic occurrence of an event.

\section{Evaluation}
\label{sec_evaluation}

\subsection{Target Scenario Development Methods and Instances}
\label{sec_evaluation_scenario_instances}

To evaluate the conceptual scenario variables (SVs) and conceptual scenario model (CSM) defined and developed in this paper, this section analyzes/evaluates five scenario development/specification methods in Automated Driving System (ADS) domain. Because the ADS domain, along with military and aviation domains, is one of the most active field of research and standardization of scenario-based engineering, scenario methods of the ADS field were selected as the application domain. Based on official documents and publications of each method, available scenario instances and conceptual data within each instance were manually extracted and inspected.

As introduced in Section~\ref{sec_related_work_ads}, the first and second methods are \textit{ASAM OpenSCENARIO} (v2.0), a standard of ADAS for testing and validation of ADSs, and \textit{PEGASUS}, a methodology of scenario-based safety analysis and testing. These methods define engineer-friendly domain-specific scenario description language for expressing highly-automated maneuver descriptions at multiple abstraction levels. Both methods also support the development of scenario execution models (i.e., test/simulation scenarios) for industrial application. Additionally, other methods proposed by B. Schutt \etal~\cite{schutt2020sceml}, J. Bach \etal~\cite{bach2016model_based}, C. M. Richard \etal~\cite{richard2006TaskAO}, and D. J. Fremont\etal~\cite{fremont2019scenic} were analyzed in terms of scenario constructs. The first four methods develop scenarios for the test development, while the fifth method aims for scenario-based task analysis and workload estimation, and the sixth method aims to propose a probabilistic programming language for scene generation.

Due to the space constraints in this paper, similar SVs centered around primary SVs were merged to identify data corresponding to 3 method-level SVs, 3 suite-level SVs, 8 scenario-level SVs, and 7 event-level SVs. Similar to the literature review, data was extracted and analyzed based on keywords from exemplary scenario instances and explanation introduced in each method's representative publication(s), as shown in Table~\ref{tab_analysis_scenario_methods}.

\subsection{Evaluation of Scenario Variables}
\label{sec_evaluation_sv}

Analysis of the SVs does not show whether a scenario development method is superior or more effective than other methods. However, it provides information on which semantic domain each method primarily focuses on and to support scenario development in terms of how the types of data differ in comparison to other methods. Because the SVs in this paper were defined by looking into the actual scenario instances, various conceptual data used in leading scenario methods were analyzed at meta-class levels. The information contained in each cell of Table~\ref{tab_analysis_scenario_methods} can be viewed as a class or instantiated data for an actual scenario specification. This section explains the observation and implications from the SV-based analysis.

\afterpage{%
    \thispagestyle{empty}%
    \centering
    \begin{landscape}%

\begin{table}[h]
\caption{Analysis of scenario specification methods in terms of primary scenario variables}
\centering
\resizebox{1.25\textheight}{!}{
}                                                                                                                                                                                                                                                                                                                                                                                                & \vcell{\textcolor[rgb]{0.149,0.149,0.149}{Scene,~}\textcolor[rgb]{0.149,0.149,0.149}{SafetyArgument,~}\textcolor[rgb]{0.149,0.149,0.149}{FailureTree,~}\textcolor[rgb]{0.149,0.149,0.149}{RegulatoryElement}\textcolor[rgb]{0.149,0.149,0.149}{,~}\textcolor[rgb]{0.149,0.149,0.149}{AcceptanceModel}\textcolor[rgb]{0.149,0.149,0.149}{,~}\textcolor[rgb]{0.149,0.149,0.149}{}\textcolor[rgb]{0.149,0.149,0.149}{RegionOfInterest}\textcolor[rgb]{0.149,0.149,0.149}{, Virtual/}\textcolor[rgb]{0.149,0.149,0.149}{RealDataSource}\textcolor[rgb]{0.149,0.149,0.149}{,~}\textcolor[rgb]{0.149,0.149,0.149}{RoadUserData}\textcolor[rgb]{0.149,0.149,0.149}{,~}\textcolor[rgb]{0.149,0.149,0.149}{AccidentModel}} & \vcell{
\begin{tabular}[b]{@{}l@{}}\textcolor[rgb]{0.149,0.149,0.149}{TrafficSequence}\textcolor[rgb]{0.149,0.149,0.149}{, }\\\textcolor[rgb]{0.149,0.149,0.149}{Setting (using }\\\textcolor[rgb]{0.149,0.149,0.149}{Functional Scenario }\\\textcolor[rgb]{0.149,0.149,0.149}{or }\textcolor[rgb]{0.149,0.149,0.149}{Misc}\textcolor[rgb]{0.149,0.149,0.149}{)}\end{tabular}} & \vcell{\begin{tabular}[b]{@{}l@{}}\textcolor[rgb]{0.149,0.149,0.149}{Scene, }\textcolor[rgb]{0.149,0.149,0.149}{RegulatoryElement}\textcolor[rgb]{0.149,0.149,0.149}{, }\\\textcolor[rgb]{0.149,0.149,0.149}{Domain Model, Rule Sets, }\textcolor[rgb]{0.149,0.149,0.149}{}\\\textcolor[rgb]{0.149,0.149,0.149}{ManeuverType}\textcolor[rgb]{0.149,0.149,0.149}{, }\textcolor[rgb]{0.149,0.149,0.149}{}\\\textcolor[rgb]{0.149,0.149,0.149}{PerceptionLayers}\end{tabular}} & \vcell{\begin{tabular}[b]{@{}l@{}}\textcolor[rgb]{0.149,0.149,0.149}{RegulatoryElement}\textcolor[rgb]{0.149,0.149,0.149}{, }\\\textcolor[rgb]{0.149,0.149,0.149}{}\textcolor[rgb]{0.149,0.149,0.149}{Historical/Operational Data}\\\textcolor[rgb]{0.149,0.149,0.149}{}\end{tabular}}                                                                                                                                            & \vcell{\begin{tabular}[b]{@{}l@{}}\textcolor[rgb]{0.149,0.149,0.149}{Scenario(Default) }\\\textcolor[rgb]{0.149,0.149,0.149}{Requirements,}\\\textcolor[rgb]{0.149,0.149,0.149}{Property,}\textcolor[rgb]{0.149,0.149,0.149}{ImagesOfCars}\end{tabular}}                                                                        \\[-\rowheight]
                                                         & \printcelltop                                                                                                                                                                       & \printcelltop                                                                                                                                                                                                                                                                                                                                                                                                                                                                                                                                                                                                                                                                                                                                                                                                                                                       & \printcelltop                                                                                                                                                                                                                                                                                                                                                                                                                                                                                                                                                                                                                                                                                                                                                                             & \printcelltop                                                                                                                                                                                                                                                                                                                                                                  & \printcelltop                                                                                                                                                                                                                                                                                                                                                                                                                                                                                                     & \printcelltop                                                                                                                                                                                                                                                                                                                                                                                                                     & \printcelltop                                                                                                                                                                                                                                                                                                                   \\ 
\cline{2-8}
                                                         & \vcell{\textcolor[rgb]{0.149,0.149,0.149}{Composition}}                                                                                                                             & \vcell{\textcolor[rgb]{0.149,0.149,0.149}{Scene-based,Causality,~}\textcolor[rgb]{0.149,0.149,0.149}{Baseline-Alternative}}                                                                                                                                                                                                                                                                                                                                                                                                                                                                                                                                                                                                                                                                                                                                         & \vcell{\textcolor[rgb]{0.149,0.149,0.149}{Ordinary-Critical}}                                                                                                                                                                                                                                                                                                                                                                                                                                                                                                                                                                                                                                                                                                                             & \vcell{\textcolor[rgb]{0.149,0.149,0.149}{Situation-based}}                                                                                                                                                                                                                                                                                                                    & \vcell{\begin{tabular}[b]{@{}l@{}}\textcolor[rgb]{0.149,0.149,0.149}{Main-Alternative,}\\\textcolor[rgb]{0.149,0.149,0.149}{Causality}\end{tabular}}                                                                                                                                                                                                                                                                                                                                                              & \vcell{\textcolor[rgb]{0.149,0.149,0.149}{ScenarioSegments}}                                                                                                                                                                                                                                                                                                                                                                      & \vcell{}                                                                                                                                                                                                                                                                                                                        \\[-\rowheight]
                                                         & \printcelltop                                                                                                                                                                       & \printcelltop                                                                                                                                                                                                                                                                                                                                                                                                                                                                                                                                                                                                                                                                                                                                                                                                                                                       & \printcelltop                                                                                                                                                                                                                                                                                                                                                                                                                                                                                                                                                                                                                                                                                                                                                                             & \printcelltop                                                                                                                                                                                                                                                                                                                                                                  & \printcelltop                                                                                                                                                                                                                                                                                                                                                                                                                                                                                                     & \printcelltop                                                                                                                                                                                                                                                                                                                                                                                                                     & \printcelltop                                                                                                                                                                                                                                                                                                                   \\ 
\hline
\multirow{8}{*}{\textcolor[rgb]{0.149,0.149,0.149}{Scn}} & \vcell{\textcolor[rgb]{0.149,0.149,0.149}{Constituent-Events}}                                                                                                                      & \vcell{\begin{tabular}[b]{@{}l@{}}\textcolor[rgb]{0.149,0.149,0.149}{Serial/Sequential,Parallel, }\\\textcolor[rgb]{0.149,0.149,0.149}{Synchronized, Event-driven, XOR}\end{tabular}}                                                                                                                                                                                                                                                                                                                                                                                                                                                                                                                                                                                                                                                                               & \vcell{\begin{tabular}[b]{@{}l@{}}\textcolor[rgb]{0.149,0.149,0.149}{BaseEvent}\textcolor[rgb]{0.149,0.149,0.149}{, }\\\textcolor[rgb]{0.149,0.149,0.149}{Faulty/}\textcolor[rgb]{0.149,0.149,0.149}{FailureEvent}\end{tabular}}                                                                                                                                                                                                                                                                                                                                                                                                                                                                                                                                                          & \vcell{\begin{tabular}[b]{@{}l@{}}\textcolor[rgb]{0.149,0.149,0.149}{EventConnection, }\\\textcolor[rgb]{0.149,0.149,0.149}{Join, Sync}\end{tabular}}                                                                                                                                                                                                                          & \vcell{\textcolor[rgb]{0.149,0.149,0.149}{Act-based,Event-driven}}                                                                                                                                                                                                                                                                                                                                                                                                                                                & \vcell{\textcolor[rgb]{0.149,0.149,0.149}{Sequential}}                                                                                                                                                                                                                                                                                                                                                                            & \vcell{}                                                                                                                                                                                                                                                                                                                        \\[-\rowheight]
                                                         & \printcelltop                                                                                                                                                                       & \printcelltop                                                                                                                                                                                                                                                                                                                                                                                                                                                                                                                                                                                                                                                                                                                                                                                                                                                       & \printcelltop                                                                                                                                                                                                                                                                                                                                                                                                                                                                                                                                                                                                                                                                                                                                                                             & \printcelltop                                                                                                                                                                                                                                                                                                                                                                  & \printcelltop                                                                                                                                                                                                                                                                                                                                                                                                                                                                                                     & \printcelltop                                                                                                                                                                                                                                                                                                                                                                                                                     & \printcelltop                                                                                                                                                                                                                                                                                                                   \\ 
\cline{2-8}
                                                         & \vcell{\textcolor[rgb]{0.149,0.149,0.149}{Target}}                                                                                                                                  & \vcell{\textcolor[rgb]{0.149,0.149,0.149}{TestDescription}}                                                                                                                                                                                                                                                                                                                                                                                                                                                                                                                                                                                                                                                                                                                                                                                                         & \vcell{\begin{tabular}[b]{@{}l@{}}\textcolor[rgb]{0.149,0.149,0.149}{Description ofAutomated }\\\textcolor[rgb]{0.149,0.149,0.149}{Driving Functions, Criticality}\end{tabular}}                                                                                                                                                                                                                                                                                                                                                                                                                                                                                                                                                                                                          & \vcell{\begin{tabular}[b]{@{}l@{}}\textcolor[rgb]{0.149,0.149,0.149}{Multi-levelBehavior}\\\textcolor[rgb]{0.149,0.149,0.149}{Description}\end{tabular}}                                                                                                                                                                                                                       & \vcell{\textcolor[rgb]{0.149,0.149,0.149}{BehaviorDescription}}                                                                                                                                                                                                                                                                                                                                                                                                                                                   & \vcell{\textcolor[rgb]{0.149,0.149,0.149}{Task~}}                                                                                                                                                                                                                                                                                                                                                                                 & \vcell{}                                                                                                                                                                                                                                                                                                                        \\[-\rowheight]
                                                         & \printcelltop                                                                                                                                                                       & \printcelltop                                                                                                                                                                                                                                                                                                                                                                                                                                                                                                                                                                                                                                                                                                                                                                                                                                                       & \printcelltop                                                                                                                                                                                                                                                                                                                                                                                                                                                                                                                                                                                                                                                                                                                                                                             & \printcelltop                                                                                                                                                                                                                                                                                                                                                                  & \printcelltop                                                                                                                                                                                                                                                                                                                                                                                                                                                                                                     & \printcelltop                                                                                                                                                                                                                                                                                                                                                                                                                     & \printcelltop                                                                                                                                                                                                                                                                                                                   \\ 
\cline{2-8}
                                                         & \vcell{\textcolor[rgb]{0.149,0.149,0.149}{Inputs}}                                                                                                                                  & \vcell{\begin{tabular}[b]{@{}l@{}}\textcolor[rgb]{0.149,0.149,0.149}{ScenarioMeta-Data, Story,}\textcolor[rgb]{0.149,0.149,0.149}{Route,~}\textcolor[rgb]{0.149,0.149,0.149}{Procedure,~}\textcolor[rgb]{0.149,0.149,0.149}{}\\\textcolor[rgb]{0.149,0.149,0.149}{Entities/}\textcolor[rgb]{0.149,0.149,0.149}{ScenarioObjects~}\textcolor[rgb]{0.149,0.149,0.149}{(Vehicles (Movable, }\\\textcolor[rgb]{0.149,0.149,0.149}{Stationary), Driver),~Parameters (Distribution) }\\\textcolor[rgb]{0.149,0.149,0.149}{File,~}\textcolor[rgb]{0.149,0.149,0.149}{Weather,~}\textcolor[rgb]{0.149,0.149,0.149}{StationaryCondition}\textcolor[rgb]{0.149,0.149,0.149}{,~}\\\textcolor[rgb]{0.149,0.149,0.149}{}\textcolor[rgb]{0.149,0.149,0.149}{ManeuverGroup}\textcolor[rgb]{0.149,0.149,0.149}{,~}\textcolor[rgb]{0.149,0.149,0.149}{CatalogReference}\end{tabular}} & \vcell{\begin{tabular}[b]{@{}l@{}}\textcolor[rgb]{0.149,0.149,0.149}{EnvironmentCondition~}\textcolor[rgb]{0.149,0.149,0.149}{(Weather), }\textcolor[rgb]{0.149,0.149,0.149}{}\\\textcolor[rgb]{0.149,0.149,0.149}{SAELevel}\textcolor[rgb]{0.149,0.149,0.149}{, Risk, Actor (Vehicle),}\\\textcolor[rgb]{0.149,0.149,0.149}{Sensors, Parameters }\\\textcolor[rgb]{0.149,0.149,0.149}{(}\textcolor[rgb]{0.149,0.149,0.149}{FrequencyDistribution}\textcolor[rgb]{0.149,0.149,0.149}{)}\end{tabular}}                                                                                                                                                                                                                                                                                     & \vcell{\begin{tabular}[b]{@{}l@{}}\textcolor[rgb]{0.149,0.149,0.149}{Weather, Actor(Wheeler, }\\\textcolor[rgb]{0.149,0.149,0.149}{Pedestrian), Parameter }\\\textcolor[rgb]{0.149,0.149,0.149}{(Range)}\end{tabular}}                                                                                                                                                         & \vcell{\begin{tabular}[b]{@{}l@{}}\textcolor[rgb]{0.149,0.149,0.149}{Scenery (RoadNetwork),}\\\textcolor[rgb]{0.149,0.149,0.149}{Participant, Route, }\\\textcolor[rgb]{0.149,0.149,0.149}{Situation, Weather}\end{tabular}}                                                                                                                                                                                                                                                                                      & \vcell{\begin{tabular}[b]{@{}l@{}}\textcolor[rgb]{0.149,0.149,0.149}{Vehicle,Scenario Segments}\textcolor[rgb]{0.149,0.149,0.149}{}\\\textcolor[rgb]{0.149,0.149,0.149}{Assumption, Workload }\\\textcolor[rgb]{0.149,0.149,0.149}{Demand (with Bottlenecks), }\\\textcolor[rgb]{0.149,0.149,0.149}{Configuration,~}\textcolor[rgb]{0.149,0.149,0.149}{Milestones,}\\\textcolor[rgb]{0.149,0.149,0.149}{Parameters}\end{tabular}} & \vcell{\begin{tabular}[b]{@{}l@{}}\textcolor[rgb]{0.149,0.149,0.149}{Objects,~}\textcolor[rgb]{0.149,0.149,0.149}{ScenarioFile}\textcolor[rgb]{0.149,0.149,0.149}{, }\\\textcolor[rgb]{0.149,0.149,0.149}{Statement,~}\textcolor[rgb]{0.149,0.149,0.149}{}\\\textcolor[rgb]{0.149,0.149,0.149}{GlobalParameters}\end{tabular}}  \\[-\rowheight]
                                                         & \printcelltop                                                                                                                                                                       & \printcelltop                                                                                                                                                                                                                                                                                                                                                                                                                                                                                                                                                                                                                                                                                                                                                                                                                                                       & \printcelltop                                                                                                                                                                                                                                                                                                                                                                                                                                                                                                                                                                                                                                                                                                                                                                             & \printcelltop                                                                                                                                                                                                                                                                                                                                                                  & \printcelltop                                                                                                                                                                                                                                                                                                                                                                                                                                                                                                     & \printcelltop                                                                                                                                                                                                                                                                                                                                                                                                                     & \printcelltop                                                                                                                                                                                                                                                                                                                   \\ 
\cline{2-8}
                                                         & \vcell{\textcolor[rgb]{0.149,0.149,0.149}{Outputs}}                                                                                                                                 & \vcell{\begin{tabular}[b]{@{}l@{}}\textcolor[rgb]{0.149,0.149,0.149}{ScenarioLibrary}\textcolor[rgb]{0.149,0.149,0.149}{/Element,~}\textcolor[rgb]{0.149,0.149,0.149}{}\\\textcolor[rgb]{0.149,0.149,0.149}{ParameterMeasurementResults}\end{tabular}}                                                                                                                                                                                                                                                                                                                                                                                                                                                                                                                                                                                                              & \vcell{\begin{tabular}[b]{@{}l@{}}\textcolor[rgb]{0.149,0.149,0.149}{TestSpecification,~}\textcolor[rgb]{0.149,0.149,0.149}{Probability and Security }\\\textcolor[rgb]{0.149,0.149,0.149}{of occurrence of (harmful) outcomes}\end{tabular}}                                                                                                                                                                                                                                                                                                                                                                                                                                                                                                                                             & \vcell{\textcolor[rgb]{0.149,0.149,0.149}{Scenario File}}                                                                                                                                                                                                                                                                                                                      & \vcell{\textcolor[rgb]{0.149,0.149,0.149}{Test Cases, Recorded Data}}                                                                                                                                                                                                                                                                                                                                                                                                                                             & \vcell{\textcolor[rgb]{0.149,0.149,0.149}{WorkloadProfile}}                                                                                                                                                                                                                                                                                                                                                                       & \vcell{}                                                                                                                                                                                                                                                                                                                        \\[-\rowheight]
                                                         & \printcelltop                                                                                                                                                                       & \printcelltop                                                                                                                                                                                                                                                                                                                                                                                                                                                                                                                                                                                                                                                                                                                                                                                                                                                       & \printcelltop                                                                                                                                                                                                                                                                                                                                                                                                                                                                                                                                                                                                                                                                                                                                                                             & \printcelltop                                                                                                                                                                                                                                                                                                                                                                  & \printcelltop                                                                                                                                                                                                                                                                                                                                                                                                                                                                                                     & \printcelltop                                                                                                                                                                                                                                                                                                                                                                                                                     & \printcelltop                                                                                                                                                                                                                                                                                                                   \\ 
\cline{2-8}
                                                         & \vcell{\textcolor[rgb]{0.149,0.149,0.149}{Condition}}                                                                                                                               & \vcell{\begin{tabular}[b]{@{}l@{}}\textcolor[rgb]{0.149,0.149,0.149}{EnvironmentCondition}\textcolor[rgb]{0.149,0.149,0.149}{,~}\textcolor[rgb]{0.149,0.149,0.149}{EntityConstraints}\textcolor[rgb]{0.149,0.149,0.149}{}\\\textcolor[rgb]{0.149,0.149,0.149}{Software/Hardware Constraints,}\textcolor[rgb]{0.149,0.149,0.149}{}\\\textcolor[rgb]{0.149,0.149,0.149}{TerminationCondition}\textcolor[rgb]{0.149,0.149,0.149}{,}\end{tabular}}                                                                                                                                                                                                                                                                                                                                                                                                                      & \vcell{\textcolor[rgb]{0.149,0.149,0.149}{EnvironmentCondition}\textcolor[rgb]{0.149,0.149,0.149}{, Traffic}}                                                                                                                                                                                                                                                                                                                                                                                                                                                                                                                                                                                                                                                                             & \vcell{}                                                                                                                                                                                                                                                                                                                                                                       & \vcell{\begin{tabular}[b]{@{}l@{}}\textcolor[rgb]{0.149,0.149,0.149}{TransitionCondition}\textcolor[rgb]{0.149,0.149,0.149}{ }\\\textcolor[rgb]{0.149,0.149,0.149}{(Situation)}\end{tabular}}                                                                                                                                                                                                                                                                                                                     & \vcell{\begin{tabular}[b]{@{}l@{}}\textcolor[rgb]{0.149,0.149,0.149}{TrafficCondition}\textcolor[rgb]{0.149,0.149,0.149}{,~}\textcolor[rgb]{0.149,0.149,0.149}{DriverCondition}\textcolor[rgb]{0.149,0.149,0.149}{,}\\\textcolor[rgb]{0.149,0.149,0.149}{TimeConstraints}\textcolor[rgb]{0.149,0.149,0.149}{,~}\textcolor[rgb]{0.149,0.149,0.149}{ResourceConstraints}\end{tabular}}                                              & \vcell{\begin{tabular}[b]{@{}l@{}}\textcolor[rgb]{0.149,0.149,0.149}{Hard/SoftConstraints,}\\\textcolor[rgb]{0.149,0.149,0.149}{Traffic/}\textcolor[rgb]{0.149,0.149,0.149}{RoadCondition}\end{tabular}}                                                                                                                        \\[-\rowheight]
                                                         & \printcelltop                                                                                                                                                                       & \printcelltop                                                                                                                                                                                                                                                                                                                                                                                                                                                                                                                                                                                                                                                                                                                                                                                                                                                       & \printcelltop                                                                                                                                                                                                                                                                                                                                                                                                                                                                                                                                                                                                                                                                                                                                                                             & \printcelltop                                                                                                                                                                                                                                                                                                                                                                  & \printcelltop                                                                                                                                                                                                                                                                                                                                                                                                                                                                                                     & \printcelltop                                                                                                                                                                                                                                                                                                                                                                                                                     & \printcelltop                                                                                                                                                                                                                                                                                                                   \\ 
\cline{2-8}
                                                         & \vcell{\textcolor[rgb]{0.149,0.149,0.149}{Temporal}}                                                                                                                                & \vcell{\textcolor[rgb]{0.149,0.149,0.149}{ScenarioTime}\textcolor[rgb]{0.149,0.149,0.149}{, Timeline}}                                                                                                                                                                                                                                                                                                                                                                                                                                                                                                                                                                                                                                                                                                                                                              & \vcell{\textcolor[rgb]{0.149,0.149,0.149}{Time-Continuity, Start/End Time}}                                                                                                                                                                                                                                                                                                                                                                                                                                                                                                                                                                                                                                                                                                               & \vcell{}                                                                                                                                                                                                                                                                                                                                                                       & \vcell{}                                                                                                                                                                                                                                                                                                                                                                                                                                                                                                          & \vcell{\begin{tabular}[b]{@{}l@{}}\textcolor[rgb]{0.149,0.149,0.149}{Timeline/}\textcolor[rgb]{0.149,0.149,0.149}{TimeFrame}\textcolor[rgb]{0.149,0.149,0.149}{,~}\textcolor[rgb]{0.149,0.149,0.149}{}\\\textcolor[rgb]{0.149,0.149,0.149}{Temporal Milestones, Interval}\end{tabular}%
}                                                                                                    & \vcell{\textcolor[rgb]{0.149,0.149,0.149}{Actor (Ego),Vision}}                                                                                                                                                                                                                                                                  \\[-\rowheight]
                                                         & \printcelltop                                                                                                                                                                       & \printcelltop                                                                                                                                                                                                                                                                                                                                                                                                                                                                                                                                                                                                                                                                                                                                                                                                                                                       & \printcelltop                                                                                                                                                                                                                                                                                                                                                                                                                                                                                                                                                                                                                                                                                                                                                                             & \printcelltop                                                                                                                                                                                                                                                                                                                                                                  & \printcelltop                                                                                                                                                                                                                                                                                                                                                                                                                                                                                                     & \printcelltop                                                                                                                                                                                                                                                                                                                                                                                                                     & \printcelltop                                                                                                                                                                                                                                                                                                                   \\ 
\cline{2-8}
                                                         & \vcell{\textcolor[rgb]{0.149,0.149,0.149}{Temporal}}                                                                                                                                & \vcell{\textcolor[rgb]{0.149,0.149,0.149}{Duration,(Relative) Time,~}\textcolor[rgb]{0.149,0.149,0.149}{TimeReference}}                                                                                                                                                                                                                                                                                                                                                                                                                                                                                                                                                                                                                                                                                                                                             & \vcell{\textcolor[rgb]{0.149,0.149,0.149}{Time}}                                                                                                                                                                                                                                                                                                                                                                                                                                                                                                                                                                                                                                                                                                                                          & \vcell{}                                                                                                                                                                                                                                                                                                                                                                       & \vcell{\textcolor[rgb]{0.149,0.149,0.149}{Time}}                                                                                                                                                                                                                                                                                                                                                                                                                                                                  & \vcell{\textcolor[rgb]{0.149,0.149,0.149}{(Relative)Timing,Duration}}                                                                                                                                                                                                                                                                                                                                                             & \vcell{}                                                                                                                                                                                                                                                                                                                        \\[-\rowheight]
                                                         & \printcelltop                                                                                                                                                                       & \printcelltop                                                                                                                                                                                                                                                                                                                                                                                                                                                                                                                                                                                                                                                                                                                                                                                                                                                       & \printcelltop                                                                                                                                                                                                                                                                                                                                                                                                                                                                                                                                                                                                                                                                                                                                                                             & \printcelltop                                                                                                                                                                                                                                                                                                                                                                  & \printcelltop                                                                                                                                                                                                                                                                                                                                                                                                                                                                                                     & \printcelltop                                                                                                                                                                                                                                                                                                                                                                                                                     & \printcelltop                                                                                                                                                                                                                                                                                                                   \\ 
\cline{2-8}
                                                         & \vcell{\textcolor[rgb]{0.149,0.149,0.149}{Geospatial}}                                                                                                                              & \vcell{\begin{tabular}[b]{@{}l@{}}\textcolor[rgb]{0.149,0.149,0.149}{(Cartesian)Coordinate, Speed }\\\textcolor[rgb]{0.149,0.149,0.149}{(Relative/Absolute), Position, }\\\textcolor[rgb]{0.149,0.149,0.149}{Direction/Angle, Distance}\end{tabular}}                                                                                                                                                                                                                                                                                                                                                                                                                                                                                                                                                                                                               & \vcell{\begin{tabular}[b]{@{}l@{}}\textcolor[rgb]{0.149,0.149,0.149}{Coordinate,Distance, Velocity/Speed, }\\\textcolor[rgb]{0.149,0.149,0.149}{Orientation}\end{tabular}}                                                                                                                                                                                                                                                                                                                                                                                                                                                                                                                                                                                                                & \vcell{\begin{tabular}[b]{@{}l@{}}\textcolor[rgb]{0.149,0.149,0.149}{Radius,Direction, }\\\textcolor[rgb]{0.149,0.149,0.149}{Velocity}\end{tabular}}                                                                                                                                                                                                                           & \vcell{\textcolor[rgb]{0.149,0.149,0.149}{Lane,Position, Distance}}                                                                                                                                                                                                                                                                                                                                                                                                                                               & \vcell{\textcolor[rgb]{0.149,0.149,0.149}{Distance,Direction, Speed}}                                                                                                                                                                                                                                                                                                                                                             & \vcell{\begin{tabular}[b]{@{}l@{}}\textcolor[rgb]{0.149,0.149,0.149}{LocalCoordinate, Offset,}\\\textcolor[rgb]{0.149,0.149,0.149}{Direction, Position, Spot, }\textcolor[rgb]{0.149,0.149,0.149}{}\\\textcolor[rgb]{0.149,0.149,0.149}{Distance}\end{tabular}}                                                                 \\[-\rowheight]
                                                         & \printcelltop                                                                                                                                                                       & \printcelltop                                                                                                                                                                                                                                                                                                                                                                                                                                                                                                                                                                                                                                                                                                                                                                                                                                                       & \printcelltop                                                                                                                                                                                                                                                                                                                                                                                                                                                                                                                                                                                                                                                                                                                                                                             & \printcelltop                                                                                                                                                                                                                                                                                                                                                                  & \printcelltop                                                                                                                                                                                                                                                                                                                                                                                                                                                                                                     & \printcelltop                                                                                                                                                                                                                                                                                                                                                                                                                     & \printcelltop                                                                                                                                                                                                                                                                                                                   \\ 
\cline{2-8}
                                                         & \vcell{\textcolor[rgb]{0.149,0.149,0.149}{Uncertainty}}                                                                                                                             & \vcell{\textcolor[rgb]{0.149,0.149,0.149}{StochasticEvent}}                                                                                                                                                                                                                                                                                                                                                                                                                                                                                                                                                                                                                                                                                                                                                                                                         & \vcell{\begin{tabular}[b]{@{}l@{}}\textcolor[rgb]{0.149,0.149,0.149}{ProbabilityOfOccurrence}\textcolor[rgb]{0.149,0.149,0.149}{, Controllability, }\\\textcolor[rgb]{0.149,0.149,0.149}{Fluctuation, Variability}\end{tabular}}                                                                                                                                                                                                                                                                                                                                                                                                                                                                                                                    & \vcell{\begin{tabular}[b]{@{}l@{}}\textcolor[rgb]{0.149,0.149,0.149}{ProbabilityDistribution}\textcolor[rgb]{0.149,0.149,0.149}{}\\\textcolor[rgb]{0.149,0.149,0.149}{(of Params)}\end{tabular}}                                                                                                                                                                               & \vcell{}                                                                                                                                                                                                                                                                                                                                                                                                                                                                                                          & \vcell{\begin{tabular}[b]{@{}l@{}}\textcolor[rgb]{0.149,0.149,0.149}{Distribution of Information }\\\textcolor[rgb]{0.149,0.149,0.149}{Perceived}\end{tabular}}                                                                                                                                                                                                                              & \vcell{\textcolor[rgb]{0.149,0.149,0.149}{ProbabilityDistribution}}                                                                                                                                                                                                                                                             \\[-\rowheight]
                                                         & \printcelltop                                                                                                                                                                       & \printcelltop                                                                                                                                                                                                                                                                                                                                                                                                                                                                                                                                                                                                                                                                                                                                                                                                                                                       & \printcellmiddle                                                                                                                                                                                                                                                                                                                                                                                                                                                                                                                                                                                                                                                                                                                                                                          & \printcellmiddle                                                                                                                                                                                                                                                                                                                                                               & \printcellmiddle                                                                                                                                                                                                                                                                                                                                                                                                                                                                                                  & \printcellmiddle                                                                                                                                                                                                                                                                                                                                                                                                                  & \printcellmiddle                                                                                                                                                                                                                                                                                                                \\
\hline

\end{tabular}
}\label{tab_analysis_scenario_methods}
\end{table}
    
    \end{landscape}
    \clearpage%
}

First, the investigated methods contain different types of data depending on the engineering activity, phase, and purpose even when applied in the same application domain. As described above, a set of data included in a scenario development method is considered as a semantic domain for scenario specification. Therefore, it is helpful in establishing the general requirements needed for appropriately employing a scenario method. For example, C. M. Richard \etal's approach provides semantics (e.g., workload demand, bottlenecks) focused on engineering activity of scenario-wide task analysis, but does not provide an actual test development. On the other hand, J. Bach \etal's approach supports visualization tools (scenario editors) to effectively support the development of test scenarios, while other methods support the specification by providing (formal) scenario description languages.

Second, central information each method describes through scenarios was roughly analyzed using the data obtained from the SVs. For example, when utilizing scene-centric scenarios (i.e., sequence of driving/traffic scenes), J. Bach \etal's method and D. J. Fremont \etal's method may be more appropriate than B. Schütt \etal's approach. Additionally, when safety-related features are explicitly included in the scenarios, PEGASUS method supports the identification of required inputs (e.g., \textit{SafetyArgument}, \textit{AcceptanceModel}, \textit{FailureTree}). Consequently, the result of SV-based analysis can be utilized to understand and analyze the specific aspect of a scenario method, and data types and target purpose required in each aspect can be confirmed.

Based on the analysis results of Table~\ref{tab_analysis_scenario_methods}, we could carry out comparative analysis of scenario semantics of the investigated studies, with respect to four levels (method, suite, scenario, and event).

\paragraph{Method-level}

Although the method-level data shared similarities, the level of abstraction supported by \textit{SpecMethod} and the observation/evaluation/validation criteria of \textit{ExecMethod} differed. Most methods in the ADS domain utilized means of generating logical or concrete scenarios from abstract/functional scenarios, which shows the necessity of supporting multiple abstraction levels for both specification and execution.

\paragraph{Suite-level}

Suite-level data could be classified according to the viewpoints, which showed significant difference in inputs of each method. The widely used viewpoints in ADS domain were ADS (i.e., ego vehicle) and traffic infrastructure. The ontologies of target vehicle, infrastructure, and environment were defined differently depending on the viewpoint. In addition, the composition method to coherently formulate multiple scenarios decided scenario classification criteria (e.g., ordinary-critical, baseline-alternative) within a scenario suite according to \textit{MethodPurpose} at method-level. Because ADSs are safety-critical systems which a malfunction of a driving function can cause catastrophic damages/loss, scenario development that engages safety argument is especially in demand. Additionally, unlike simple programs defined at a functional or logical levels, the suite-level inputs on physical setting, environment, and regulatory artifacts were usually required. When data-driven approach is supported (PEGASUS Method, C. M. Richard \etal's), historical data related to driving function and ADS users are also often required. Compared to the first five methods, \textit{Scenic}, proposed by D. J. Fremont \etal, is more specialized to develop realistic scenes and images of the scenes, thus it requires more specific requirements and data to create the scenes.

\paragraph{Scenario-level}

Based on the scenario-level inputs of ADS scenarios, the detailed specification of environmental context each scenario faces (or interacts with) were crucial compared to other software domains. All the analyzed methods utilized the environmental and operational context such as \textit{weatherCondition} and \textit{stationaryCondition} to process the runtime context of each scenario. These data are used to define scenario-level condition (e.g., \textit{initialCondition}, \textit{terminalCondition}) and constraints (e.g., \textit{hardware/software Constraints}, \textit{regulatoryConstraints}).

The most central data at the scenario level is provided by parameterization. Logical scenario utilizes parameter range to specify the scenario dynamics, and concrete scenario provides a set of concrete parameter values within the range. To extract more than one executable and realistic concrete scenarios from the logical scenario, the (frequency) distribution of each parameter value also can be provided. In addition, the parameter is delivered to more than one event for defining the overall probability of occurrence and event transitions.

Additionally, due to the characteristic of ADS scenarios which defines the dynamics of participants according to the flow of time within a region, temporal and geospatial abstraction are considered as important data. Based on the target timeframe/timeline of each scenario, temporal abstraction is detailed through phases, stages, or milestones. Geospatial abstraction mainly details a road network (e.g., layout, geometry) and geospatial changes to effectively represent scenario-level participants' movements.

\paragraph{Event-level}

Event-level data mostly contained similar semantic domain, including behavioral, temporal, spatial, and uncertainty properties. According to the scenario specification method, the unit behavior of an event may differ, which are often act, action (e.g., maneuver), communication (e.g., interrupt, stimuli), and activity. For not only the intended execution of each behavior, but also the representation of uncertainty and non-determinism of event occurrence, probability is frequently specified. Because an event is a unit accessing actual states or data of scenario participants (i.e., entities), the semantic domain of temporal and spatial data to capture the information on when/where to occur varied depending on the method.

\section{Discussion}
\label{sec_evaluation_discussion}

\subsection{Threats to Validity}
\label{sec_evaluation_discussion_ttv}

\paragraph{Content and Construct Validity} \textit{Content validity} involves the systematic examination of the survey contents to determine whether they cover a representative sample of the domain to be measured. \textit{Construct validity} generally refers to the degree to which a survey is legitimately conducted (i.e., legitimate experimental setup) and the survey measures what it says it measures. In other words, the content validity describes whether a systematic investigation was conducted to collect the representative data, and the construct validity describes whether the survey was conducted with a proper and correct way.

To satisfy the content validity, this study designed search keywords and queries to software and systems engineering domain, which has a high potential for employing scenario methods. However, because the selection of more than 50K publications was not feasible, the search was limited to event-based methods and target engineering activities, such as requirements engineering, design, simulation, and testing. Since these activities (and their engineering domains) were discovered as frequent use cases of scenario methods through our preliminary investigation, our data collection was expected to collect representative data of the overall contents related to scenarios and scenario methods.

Although various survey approaches were considered, this study designed a semi-systematic literature review to primarily satisfy the construct validity. Still, there are three elements that can hider this validity. First, 100 publications were finally selected based on the subjective selection criteria (\textit{relevance-criteria} of Table~\ref{tab_selection_rounds}) to conduct a full-read review. Since in-depth analysis and manual collection of conceptual data on all of the finally-selected publications require several hours of tedious work, 100 publications were further selected from the 354 publications to conduct a questionnaire-based investigation (i.e., full-read). Second, a keyword-based search was conducted on the textual description of each publication investigated. In other words, because we had to find the target data using a keyword in PDF files (e.g., we searched ``\textsf{cond*}'' for finding \textit{pre-/post-conditions}), the data could not be collected if an investigated paper did not explicitly write the term to be searched. Third, manual inspection of conceptual data was conducted in the full-read step. However, we designed and used a questionnaire and look-up tables to minimize the missed data and to maximize the consistency of the data collection.

\paragraph{Internal and External Validity} \textit{Internal validity} refers to an inductive estimate of the degree to which conclusions about causal relationships can be made, based on the measures, research setting, and whole research design (i.e., subjectiveness). \textit{External validity} refers to the extent to which the internally valid results of a study can be held to be true for other cases (i.e., generalizability). In other words, the internal validity determines whether the rationale of the internal results (e.g., scenario variables in this study) can be provided from the collected survey data, and the external validity determines whether the generalization of this study (i.e., extension to other studies, cases, and domains) is possible.

First, this study aimed to satisfy the validities by defining the characteristics and concepts from the data consolidation and classification, excluding the authors' subjective definitions. As discussed above, we reviewed the publications not only from a technical aspect, but also based on scenario descriptions developed and used in actual approaches (e.g., scenario instances). The conceptual model, an abstraction of the entire collected data set, is expected to include comprehensive results from the scenario investigation separate from the settings constructed in this study. In other words, all of the conceptual variables defined in this study are based on the concepts, terms, and data from the survey, and thus satisfying the internal validity.

Also, the investigated scenario methods were employed for both academic purposes and industrial/practical uses. Consequently, the scenario variables of our study can be considered generic concepts for a variety of engineering and application domains. In the case of domain-specific methods, the survey did not collect the entire domain-specific elements; instead they are generalized and mapped to higher-level concepts during the data collection. In addition, even though the variables are defined as generic concepts, we evaluated the applicability and extensibility of the variables using specialized scenario instances of a particular domain (ADS scenarios in this study). Therefore, the scenario variables and conceptual scenario model we developed have external validity as they can be extended and applied to external cases.

\subsection{Remaining Challenges}
\label{sec_evaluation_discussion_challenges}

The \textit{RQ3} of our survey needs to be answered by identifying challenges ahead in developing and applying scenario methods. Even though we systematically collected and conceptualized the SVs, practical/industrial application of the scenario methods remains several issues.

\paragraph{Inaccessible information of black-box participants}
Most of the investigated scenarios are developed under the assumption that the information of the scenario participants who are included or exhibit behaviors in a scene or an event is accessible. Scenario exercise is an execution stage of the specified scenario with the actual context input. At this time, there may be a limitation or restriction on data access that were not considered in the scenario planning/development stage. As the scale and the complexity of systems increase (e.g., System-of-Systems), there are more cases where the internal property and data of the constituent systems (or subsystems) are not directly accessible or controllable. To resolve this, many existing research defines various abstraction levels (e.g., function/abstract - logical - concrete/executable, in order of concreteness) and supports the specification of the parameter ranges of the property and data at the logical level. To define \textit{EventBehavioral} and \textit{EventAction} of the event-level SVs, analysis on the accessible data need to be performed. When there is a lack of information, \textit{EventUncertainty} based on a cause need to be correctly and reasonably defined.

\paragraph{Composition of multiple scenarios as a suite}
Although most scenario methods provided specific semantics and syntax to specify a scenario, these methods do not discuss how to formulate and integrate multiple scenarios into a scenario suite. As defined in the \textit{SuiteScenarioComposition} of suite-level SV, multiple scenarios in a scenario suite sharing a same viewpoint may have inter-scenario relationships, such as association (general interaction), dependency, concurrency, and causality. In addition, an event included in a scenario can depend or interact with an event of a different scenario, which can affect the scenario composition and future exercises. Ultimately, a developed scenario suite should be coherently organized. Multiple scenarios in a suite should depict the occurrence such that a scenario engineer can strategically and effectively utilize scenarios (e.g., prioritizing scenario groups in a suite).

\paragraph{Scenario-level assessment of test/simulation complexity}
As mentioned above, scenarios can be a useful communication tool throughout the software/system development. However, the complexity of functional or operational scenarios---which are written in a narrative style---from goal/requirements engineering stage and executable/concrete scenarios for test/simulation many significantly differ. For example, to describe an executable real-world-like situation in aviation or ADS scenarios, real sensor data and actuation mechanisms need to be described in detail~\cite{Schnelle2019ReviewOS}. Based on our investigation, most scenarios developed at the system analysis and architecture stage do not accompany an analysis on the cost at the execution stage. Therefore, a standardized method to evaluate the complexity of executing the developed scenario suite and scenarios, and an appropriate development process need to be provided to resolve when the complexity is above a certain threshold (e.g., decomposing to multiple sub-scenarios for upscaling in the future).

\paragraph{Top-down and bottom-up approaches for scenario development}
Scenario itself is an important artifact that needs to be complete and error-free similar to software code. However, detailed analysis and scenario development of countless contextual alternatives may cost as much as an actual system development. Although \textit{SpecMethod} and \textit{ExecMethod} of method-level SV were defined to analyze how to develop/specify and execute a scenario, most studies besides the investigated standards (e.g., \textit{OpenSCENARIO/OpenDRIVE} and \textit{PEGASUS Method}) do not specifically scrutinize the scenario development process. Except for method-level of the 4 levels, suite-scenario-event levels have a hierarchical structure where a higher-level contains lower-level elements, and thus top-down or bottom-up approach may be selected depending on where to start the development process.

\paragraph{Agile development of scenarios}
For smaller scale software (e.g., mobile app), agile development process has been suggested and actively employed to react to frequent changes and to increase transparency in developing use cases and usage scenarios. However, process for developing scenarios of large-scale complex systems (e.g., ADS incident scenario) is usually unidirectional, hierarchical, and data-centered. These characteristics challenge repetitive refinement needed to respond to goal/requirement changes and new contextual information. In future research, study on agile scenario development and support for existing methods are needed to reduce cost and increase efficiency in scenario development and maintenance. In particular, for co-simulation where multiple components (or subsystems) are modeled and simulated in distributed environment, the process to improve scenarios based on the feedback from testing/simulation results is especially required.

\section{Conclusion}
\label{sec_conclusion}

This study stemmed from the lack of conceptual basis in existing studies that suggested or utilized scenario methods. For this reason, many scenario engineers, such as methodologists and developers, have encountered some challenges and difficulties when employing the methods without well-established understanding. To resolve this issue, this work emphasizes the necessity of a conceptual framework for scenario methods and conducts a study as part of developing a conceptual scenario framework. First, to provide comprehensive understanding of scenarios and scenario methods, this study conducted a semi-systematic literature review, which investigates and collects conceptual \textit{scenario variables (SVs)} used in diverse scenario methods. Second, the collected variables (29 primary SVs, and 91 subordinate SVs) were conceptualized and organized as a \textit{conceptual scenario model (CSM)}, by meta-modeling 4 level (\textit{method-level}, \textit{suite-level}, \textit{scenario-level}, and \textit{event-level}) constructs of scenario methods. To evaluate the applicability of the SVs and CSM, three representative scenario instances were analyzed with respect to construct variables of CSM.

The ultimate goal of this study was to develop a conceptual scenario framework that can be utilized in various engineering activities and application domains. Based on the SVs and CSM, the framework will provide a scenario modeling method including a modeling language to promote the application of the scenario methods. Also, by defining dimensions (and facets), it is expected that dimension-based analysis and positioning of scenario methods are enabled by the framework, instead of analyzing the entire set of variables. Lastly, as discussed in Section~\ref{sec_evaluation_discussion}, there are remaining challenges for scenario engineers to practically employ the scenario methods, especially for actual engineering activities, such as simulation, testing, and verification. Our future work will concentrate on the framework's practical capability to support the scenario-based engineering in a particular application domain (i.e., domain-specific).

\bibliographystyle{ACM-Reference-Format}
\bibliography{bibliography}

\appendix

\onecolumn

\section{Finally Selected 100 Publications for the Full-Read}
\label{appendix_a}

\begin{longtable}[l]{lllll}
 \caption{FSP-100}\\

 \hline
  \vcell{No}    & \vcell{F. Authors}    & \vcell{Year}  & \vcell{Search Engine} & \vcell{Title}                                                                                                                                                                           \\[-\rowheight]
\printcelltop & \printcelltop         & \printcelltop & \printcelltop         & \printcelltop                                                                                                                                                                           \\ 
 \hline
 \endfirsthead

 \hline
  \vcell{No}    & \vcell{F. Authors}    & \vcell{Year}  & \vcell{Search Engine} & \vcell{Title}                                                                                                                                                                           \\[-\rowheight]
\printcelltop & \printcelltop         & \printcelltop & \printcelltop         & \printcelltop                                                                                                                                                                           \\ 
 \hline
 \endhead

 \hline
 \endfoot

 \hline
 \endlastfoot

\hline
\vcell{1}     & \vcell{Lee J.}        & \vcell{2001}  & \vcell{Scopus}        & \vcell{Verifying scenarios with time Petri-nets}                                                                                                                                        \\[-\rowheight]
\printcelltop & \printcelltop         & \printcelltop & \printcelltop         & \printcelltop                                                                                                                                                                           \\ 
\hline
\vcell{2}     & \vcell{Ohnishi}       & \vcell{2002}  & \vcell{WoS}           & \vcell{The consistency management of scenarios from different viewpoints}                                                                                                               \\[-\rowheight]
\printcelltop & \printcelltop         & \printcelltop & \printcelltop         & \printcelltop                                                                                                                                                                           \\ 
\hline
\vcell{3}     & \vcell{Nakajima T.}   & \vcell{2003}  & \vcell{Scopus}        & \vcell{\begin{tabular}[b]{@{}l@{}}Automatic Testing of Embedded Software Based on State-Transition Requirement \\Specifications\end{tabular}}                                           \\[-\rowheight]
\printcelltop & \printcelltop         & \printcelltop & \printcelltop         & \printcelltop                                                                                                                                                                           \\ 
\hline
\vcell{4}     & \vcell{Alfonso A.}    & \vcell{2004}  & \vcell{Scopus}        & \vcell{Visual timed event scenarios}                                                                                                                                                    \\
\hline
 \end{longtable}

\newpage
\onecolumn
\section{Definition of Scenario Variables}
\label{appendix_b}

%
 \twocolumn

\end{document}